\documentclass[12pt]{article}
\usepackage{mathrsfs}
\usepackage{amsmath,amssymb}
\usepackage{epsfig}
\usepackage{relsize}
\usepackage{bbm}
\usepackage{graphicx}
\usepackage{float}
\usepackage{color}
\usepackage{enumitem}

\def\({\left(}
\def\){\right)}

\def\sl(2){\alg{sl}(2)}

\def\be{\begin{equation}}
\def\ee{\end{equation}}

\newcommand{\bea}{\begin{eqnarray}}
\newcommand{\eea}{\end{eqnarray}}

\def\la{\label}

\def\ov{\over}

\newcommand{\alg}[1]{\mathfrak{#1}}
\newcommand{\su}{\alg{su}}

\newcommand{\AdS}{{\rm  AdS}_5\times {\rm S}^5}

\newcommand{\bem}{\left (\begin{matrix}}
\newcommand{\eem}{\end{matrix} \right )}

\def\lam{\lambda}

\usepackage{hyperref}
\hypersetup{
plainpages=false
}

\begin{document}

\null\vskip-24pt\hfill {\tt
TCDMATH 12-01} \vskip-1pt \hfill   {\tt HMI-12-01} \vskip0.2truecm
\begin{center}
\begin{center}
\vskip 0.8truecm {\Large\bf Scaling dimensions from the mirror TBA}
\end{center}

\renewcommand{\thefootnote}{\fnsymbol{footnote}}

\vskip 0.9truecm

Sergey Frolov\footnote{Email: frolovs@maths.tcd.ie} {}\footnote{Correspondent fellow at Steklov
Mathematical Institute, Moscow.}
 \\
\vskip 0.5cm

  {\it Hamilton Mathematics Institute and School of Mathematics, \\
~~Trinity College, Dublin 2, Ireland}

\end{center}
\vskip 1cm \noindent\centerline{\bf Abstract} \vskip 0.2cm

The mirror TBA equations  proposed by Arutyunov, Suzuki and the author are solved numerically up to 't Hooft's coupling $\lambda\approx 2340$ for several two-particle states dual to ${\cal N}=4$ SYM operators from the $\sl(2)$ sector. The data obtained for states with mode numbers $n=1,2,3,4$ is used to propose a general charge $J$  dependent  formula for the first nonvanishing subleading coefficient in the strong coupling expansion of scaling dimensions. In addition we find that the first critical and subcritical values of the coupling  for the $J=4, n=1$ operator are at $\lambda\approx 133$ and $\lambda\approx 190$, respectively.

\newpage

\tableofcontents

\renewcommand{\thefootnote}{\arabic{footnote}}
\setcounter{footnote}{0}


\section{Introduction and Summary}

The mirror Thermodynamic Bethe Ansatz (TBA) is an efficient tool to analyze 
energies of the light-cone  $\AdS$ string states, and 
through  the AdS/CFT correspondence \cite{M}, 
scaling dimensions of dual primary operators in planar ${\cal N}=4$ SYM. 
The mirror TBA originally proposed to determine the finite-size spectrum of two-dimensional relativistic integrable models \cite{Zamolodchikov90,DT96,BLZe}
reformulates the spectral problem in terms of 
thermodynamics of a so-called mirror theory
\cite{AF07} related to the original model by a double-Wick rotation.  The  $\AdS$  mirror model was studied in detail in \cite{AF07} where, in particular, the bound state spectrum and the mirror form of the Bethe-Yang (BY) equations of \cite{BS}, necessary to formulate the string hypothesis for the $\AdS$ mirror model  \cite{AF09a},  were determined. This  opened a way to derive the ground state TBA equations \cite{AF09b,BFT}, 
and to construct  excited state equations
for string states with real  \cite{GKKV09}-\cite{Sfondrini:2011rr} and complex momenta \cite{AFT11}. This was done  by using the contour deformation trick,
a procedure inspired by \cite{DT96,BLZe,KP},  and further developed in \cite{AFS09,Arutyunov:2011uz}. The TBA equations were used to analyze various aspects of the finite-size spectrum. 
It was shown in   \cite{Gromov09a} that at the large  't Hooft coupling $\lam$ energies of  semi-classical string states found by using the mirror TBA agree with explicit string theory calculations. The scaling dimension of the Konishi operator was determined up to five loops and shown
\cite{AFS10,BH10a,BH10b} to agree with L\"uscher's corrections \cite{BJ08}-\cite{Janik:2010kd}
and at four loops with explicit field-theoretic computations \cite{Sieg,Velizhanin:2008jd}.
The TBA equations for  the Konishi operator  were also solved numerically for intermediate values of the
coupling \cite{GKV09b,F10} and the results obtained agree  with various string theory considerations
  \cite {Gromov:2011de}-\cite{Beccaria:2011uz}. 

\medskip

In this paper  the mirror TBA equations of \cite{AFS09} are solved numerically for several two-particle states dual to ${\cal N}=4$ SYM operators  from the $\sl(2)$ sector with various values of the charge $J$ and mode number $n$. 
 For operators with $n\ge 2$ the scaling dimensions  are found 
up to $\lambda\approx 2340$. The same code as in  \cite{F10} is used here with minor modifications for the $J=4,n=1$ state beyond its (sub)critical value.
The mode number $n$ of a primary operator coincides with the string level of the dual string state \cite{AFS},
and at 
large values of  $\lam$ one can expand the energy of the state in an asymptotic series in powers of $1/ \sqrt[4]{n^2{\lam}} $
\be
\la{asexp0}
E_{(J,n)}(\lam)=c_{-1}\sqrt[4]{n^2{\lam}}+c_0+\frac{c_1}{
   \sqrt[4]{n^2{\lam}}}+\frac{c_2}{\sqrt{n^2{\lam}}}+\frac{c_3}{(n^2\lam)^{3/4}}+\frac{c_4}{n^2\lam}+\frac{c_5}{(n^2\lam)^{5/4}}+ \cdots\,,
   \ee
where the coefficients $c_i$ are in general nontrivial functions of $J$ and $n$.
The coefficient $c_{-1}$ of the leading term  should be in fact equal to 2 as follows from the spectrum of string theory in flat space \cite{GKV02} and asymptotic Bethe ansatz considerations \cite{AFS}.  The constant term $c_0$ is believed to vanish for two-particle states because it does in the free fermion model \cite{AF05}  describing the $\su(1|1)$ sector in the semi-classical approximation and one does not expect getting quantum corrections to $c_0$ \cite{Plefka}.  
The subleading coefficients $c_{2k-1}$ are supposed to have the following structure \cite{AF05}
\be
\la{c2km1}
c_{2k-1}=-\frac{\left(-\frac{1}{4}\right)^k \Gamma \left(k-\frac{1}{2}\right)}{\sqrt{\pi } k!}J^{2k} + b_{2k-1}(J,n)\,,
\ee
where the $J^{2k}$ term is fixed by the flat space spectrum, and $b_{2k-1}$ is a polynomial of degree $2k-1$ (or less) in $J$ with $n$ dependent coefficients. 
In particular the first nonvanishing subleading coefficient $c_1$ is  of the form $c_1=J^2/4 + b_1(J,n)$ where $b_1$ may be a linear function of $J$. 
The mirror TBA prediction for the Konishi state with $J=2, n=1$ is $b_1(2,1)=1$  \cite{GKV09b}, and it agrees  with (incompletely justified) string  computations
  \cite {Gromov:2011de}-\cite{Beccaria:2011uz}. 
In addition
in the free fermion model one finds that $b_1$ is independent of both $J$ and $n$ and equal to $1/2$ \cite{AF05}. The formulae derived in the framework of the free fermion model definitely get quantum corrections but assuming that the $J$ dependence of $b_1$ remains unchanged one immediately concludes that for any $J$ one should have $b_1(J,1)=1$. Our data for the $J=3,n=1$ and $J=4,n=1$ states indeed confirms the conclusion. Fitting the data for the $J=4,n=2$ and $J=5,n=2$ states we find that the formula also works fine: $b_1(J,2)=1$. It is tempting to assume that the same formula might be valid for any $(J,n)$ state. The analysis of the $J=6,n=3$ and $J=7,n=3$ states shows however that for $n=3$ the formula is different:
$b_1(J,3)=0$. Thus, the coefficient $c_1$ has a nontrivial $n$ dependence. 
Assuming that $b_1(J,n)$ is independent of $J$ and fitting the data for the $J=8,n=4$ state we find $b_1(J,4)=-2$. All these values of $b_1$ can be obtained from one simple formula $b_1(J,n)=n(3-n)/2$, and therefore 
\be
\la{c1}
c_{1}={J^{2}\ov 4} + {n(3-n)\ov 2}\,,\quad n=1,2,3,4\,.
\ee
It is clear that only an analytic derivation of $c_1$ can determine if this formula is valid for any $J$ and $n$. 

Coming back to the series \eqref{asexp0}, it was argued in \cite{RT} that the coefficient $c_2$ should vanish due to the high degree of supersymmetry of the model, and our data supports this.   Recently by using a conjecture from \cite{Basso}  a formula for the coefficient $c_3$ was proposed in \cite{Gr11}. Our data for $n=1$ and $n=2$ states agrees with the formula.  It would be interesting to see if there is a simple modification of the formula for higher  values of $n$. Nothing is known about other coefficients in \eqref{asexp0}. 
It is even unclear if the series expansion  is in powers of $1/\sqrt[4]{{\lam}}$ as follows from the free fermion model  \cite{AF05},  or up to an overall factor of $\sqrt[4]{{\lam}}$ it is in powers of $1/\sqrt{{\lam}}$, that is $c_{2k}=0$, as was assumed in \cite{GKV09b}. 
If it is in powers of $1/\sqrt{{\lam}}$ then this would imply that quantum corrections to the free fermion model expressions 
drastically change the structure of the strong coupling expansion. The precision of our computation is insufficient to come to a definite conclusion.  Nevertheless using the formula of \cite{Gr11} and our data we find some evidence in favor of vanishing  $c_4$ for the $n=1$ states.

\medskip

It is known \cite{AFS09} that two-particle states  from the $\sl(2)$ sector are divided into infinitely-many classes which differ by analytic properties of exact Y-functions and therefore by driving terms in the TBA equations.  The analytic properties of Y-functions depend on the coupling and at critical values of $\lam$ a state moves from 
one class to another one.
At weak coupling all the states we analyzed belong to the simplest Konishi-like class. 
The states with $n\ge2$ remain in the class up to the largest value of $\lam$ the TBA equations were solved. 
The $J=3,n=1$ and $J=4,n=1$ states however have first critical values at $\lambda\approx 950$ and $\lambda\approx 133$, respectively. The values were obtained by interpolating the data because
the iterations stopped to converge for $\lam$'s close to the critical values.
For the  $J=3,n=1$  state we could solve the equations only up to  $\lambda\approx 540$ which is pretty far from its critical value. For the  $J=4,n=1$  state the equations were solved up to $\lambda\approx 105$, and then  in accordance with \cite{AFS09} we changed the TBA equations,  jumped beyond the subcritical value to $\lambda\approx 191$ and resumed the iterations. The  iterations however stopped to converge at $\lam\approx 483$, and we do not really understand a reason. 

\medskip

The paper is organized as follows. In the next section we present
the analysis of the results of the numerical solution of the TBA equations for the states.  In section 3 our data for the states is collected.

\section{Fitting the numerical data}

\subsection{$J=3,n=1$ operator}

In the table \eqref{E31data} we present the data for the $J=3,n=1$ state. Since we could solve the TBA equations only up to $g=3.7$ to fit the data we should make assumptions about the structure of the strong coupling expansion. Assuming that 
$c_{-1}=2$ and $c_0=0$ and fitting the data in the interval $g\in [g_0, 3.7]$ we get 
{\smaller 
\bea\la{FitEJ3n1a}
\begin{array}{|c|c|c|}
\hline
g_0&\lam_0& {\rm Fit} \\\hline
1.2 & 56.8489 & 2 \sqrt[4]{\lambda }+\frac{3.1943}{\sqrt[4]{\lambda }}+\frac{0.831811}{\sqrt{\lambda
   }}-\frac{6.99044}{\lambda ^{3/4}}+\frac{13.106}{\lambda }-\frac{2.08406}{\lambda ^{5/4}} \\
 1.3 & 66.7185 & 2 \sqrt[4]{\lambda }+\frac{3.21575}{\sqrt[4]{\lambda }}+\frac{0.508957}{\sqrt{\lambda
   }}-\frac{5.18593}{\lambda ^{3/4}}+\frac{8.66792}{\lambda }+\frac{1.96869}{\lambda ^{5/4}} \\
 1.4 & 77.3777 & 2 \sqrt[4]{\lambda }+\frac{3.21158}{\sqrt[4]{\lambda }}+\frac{0.572881}{\sqrt{\lambda
   }}-\frac{5.55021}{\lambda ^{3/4}}+\frac{9.58264}{\lambda }+\frac{1.11474}{\lambda ^{5/4}} \\
 1.5 & 88.8264 & 2 \sqrt[4]{\lambda }+\frac{3.16939}{\sqrt[4]{\lambda }}+\frac{1.23048}{\sqrt{\lambda
   }}-\frac{9.36522}{\lambda ^{3/4}}+\frac{19.3462}{\lambda }-\frac{8.1862}{\lambda ^{5/4}}
    \\\hline
   \end{array}~~~~
\eea
}For the $J=3,n=1$ state we expect $c_1=J^2/4 + 1 = 3.25$ and indeed the values we get from the fitting are very close to it. Fixing then  $c_1= 3.25$ one gets
{\smaller 
\bea\la{FitEJ3n1b}
\begin{array}{|c|c|c|}
\hline
g_0&\lam_0& {\rm Fit} \\\hline
 1.2 & 56.8489 & 2 \sqrt[4]{\lambda }+\frac{3.25}{\sqrt[4]{\lambda }}+\frac{0.0297414}{\sqrt{\lambda
   }}-\frac{2.71216}{\lambda ^{3/4}}+\frac{3.08535}{\lambda }+\frac{6.61514}{\lambda ^{5/4}} \\
 1.3 & 66.7185 & 2 \sqrt[4]{\lambda }+\frac{3.25}{\sqrt[4]{\lambda }}+\frac{0.00509108}{\sqrt{\lambda
   }}-\frac{2.43564}{\lambda ^{3/4}}+\frac{2.06532}{\lambda }+\frac{7.85264}{\lambda ^{5/4}} \\
 1.4 & 77.3777 & 2 \sqrt[4]{\lambda }+\frac{3.25}{\sqrt[4]{\lambda }}-\frac{0.00358391}{\sqrt{\lambda
   }}-\frac{2.33641}{\lambda ^{3/4}}+\frac{1.6914}{\lambda }+\frac{8.31688}{\lambda ^{5/4}} \\
 1.5 & 88.8264 & 2 \sqrt[4]{\lambda }+\frac{3.25}{\sqrt[4]{\lambda }}-\frac{0.00135757}{\sqrt{\lambda
   }}-\frac{2.36233}{\lambda ^{3/4}}+\frac{1.79102}{\lambda }+\frac{8.19057}{\lambda ^{5/4}}
    \\\hline
   \end{array}~~~~
\eea
}We see that the coefficient $c_2$ becomes very small and we set it to 0: $c_2=0$
{\smaller 
\bea\la{FitEJ3n1c}
\begin{array}{|c|c|c|}
\hline
g_0&\lam_0& {\rm Fit} \\\hline
 1.2 & 56.8489 & 2 \sqrt[4]{\lambda }+\frac{3.25}{\sqrt[4]{\lambda }}-\frac{2.39691}{\lambda
   ^{3/4}}+\frac{1.99012}{\lambda }+\frac{7.86361}{\lambda ^{5/4}} \\
 1.3 & 66.7185 & 2 \sqrt[4]{\lambda }+\frac{3.25}{\sqrt[4]{\lambda }}-\frac{2.38036}{\lambda
   ^{3/4}}+\frac{1.8682}{\lambda }+\frac{8.08376}{\lambda ^{5/4}} \\
 1.4 & 77.3777 & 2 \sqrt[4]{\lambda }+\frac{3.25}{\sqrt[4]{\lambda }}-\frac{2.37618}{\lambda
   ^{3/4}}+\frac{1.83666}{\lambda }+\frac{8.14215}{\lambda ^{5/4}} \\
 1.5 & 88.8264 & 2 \sqrt[4]{\lambda }+\frac{3.25}{\sqrt[4]{\lambda }}-\frac{2.37771}{\lambda
   ^{3/4}}+\frac{1.8484}{\lambda }+\frac{8.11991}{\lambda ^{5/4}}
    \\\hline
   \end{array}~~~~
\eea
}According to the conjecture of \cite{Gr11} the coefficient $c_3$ for the states with $n=1$ is equal to
\be\la{c3}
c_3=-\frac{J^4}{64}+\frac{3 J^2}{8}-3 \zeta (3)-\frac{3}{4}\,,
\ee
and therefore for $J=3$, $c_3\approx -2.2468$. We see that the number is indeed very close to the one we get from the fit. Fixing the coefficient to this value and adding more terms to the expansion, one gets
{\smaller 
\bea\la{FitEJ3n1d}
\begin{array}{|c|c|c|}
\hline
g_0&\lam_0& {\rm Fit} \\\hline
 1.2 & 56.8489 & 2 \sqrt[4]{\lambda }+\frac{3.25}{\sqrt[4]{\lambda }}-\frac{2.2468}{\lambda
   ^{3/4}}+\frac{0.363256}{\lambda }+\frac{13.8122}{\lambda ^{5/4}}-\frac{7.89879}{\lambda ^{3/2}}+\frac{1.96167}{\lambda
   ^{7/4}} \\
 1.3 & 66.7185 & 2 \sqrt[4]{\lambda }+\frac{3.25}{\sqrt[4]{\lambda }}-\frac{2.2468}{\lambda
   ^{3/4}}+\frac{0.0493902}{\lambda }+\frac{17.2428}{\lambda ^{5/4}}-\frac{20.231}{\lambda ^{3/2}}+\frac{16.5518}{\lambda
   ^{7/4}} \\
 1.4 & 77.3777 & 2 \sqrt[4]{\lambda }+\frac{3.25}{\sqrt[4]{\lambda }}-\frac{2.2468}{\lambda
   ^{3/4}}-\frac{0.0985827}{\lambda }+\frac{18.8979}{\lambda ^{5/4}}-\frac{26.3304}{\lambda
   ^{3/2}}+\frac{23.9612}{\lambda ^{7/4}} \\
 1.5 & 88.8264 & 2 \sqrt[4]{\lambda }+\frac{3.25}{\sqrt[4]{\lambda }}-\frac{2.2468}{\lambda
   ^{3/4}}-\frac{0.0771042}{\lambda }+\frac{18.6526}{\lambda ^{5/4}}-\frac{25.4057}{\lambda
   ^{3/2}}+\frac{22.8107}{\lambda ^{7/4}}
    \\\hline
   \end{array}~~~~
\eea
}\begin{figure}[t]
\begin{center}
\includegraphics*[width=0.5\textwidth]{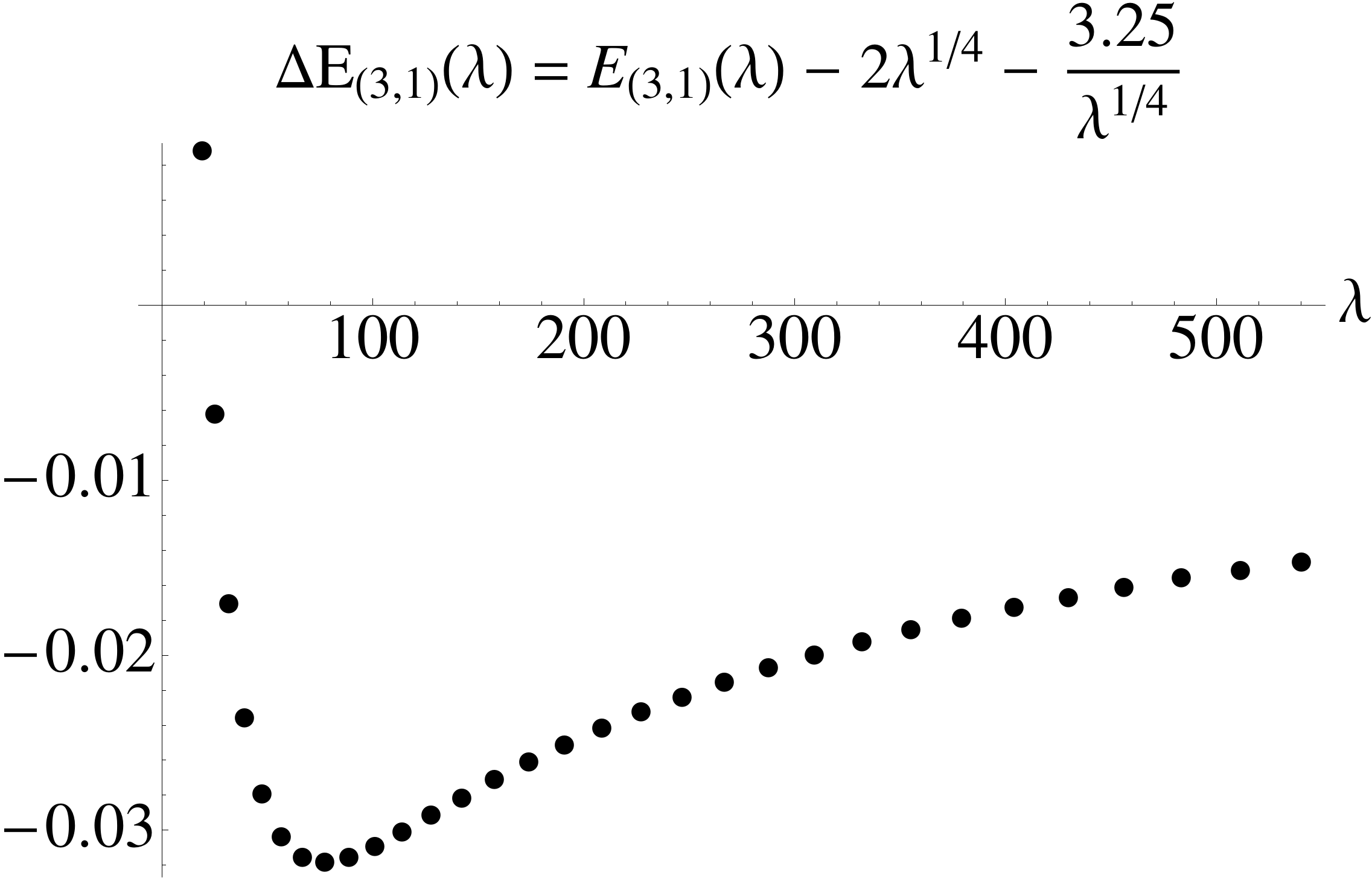}\qquad \includegraphics*[width=0.42\textwidth]{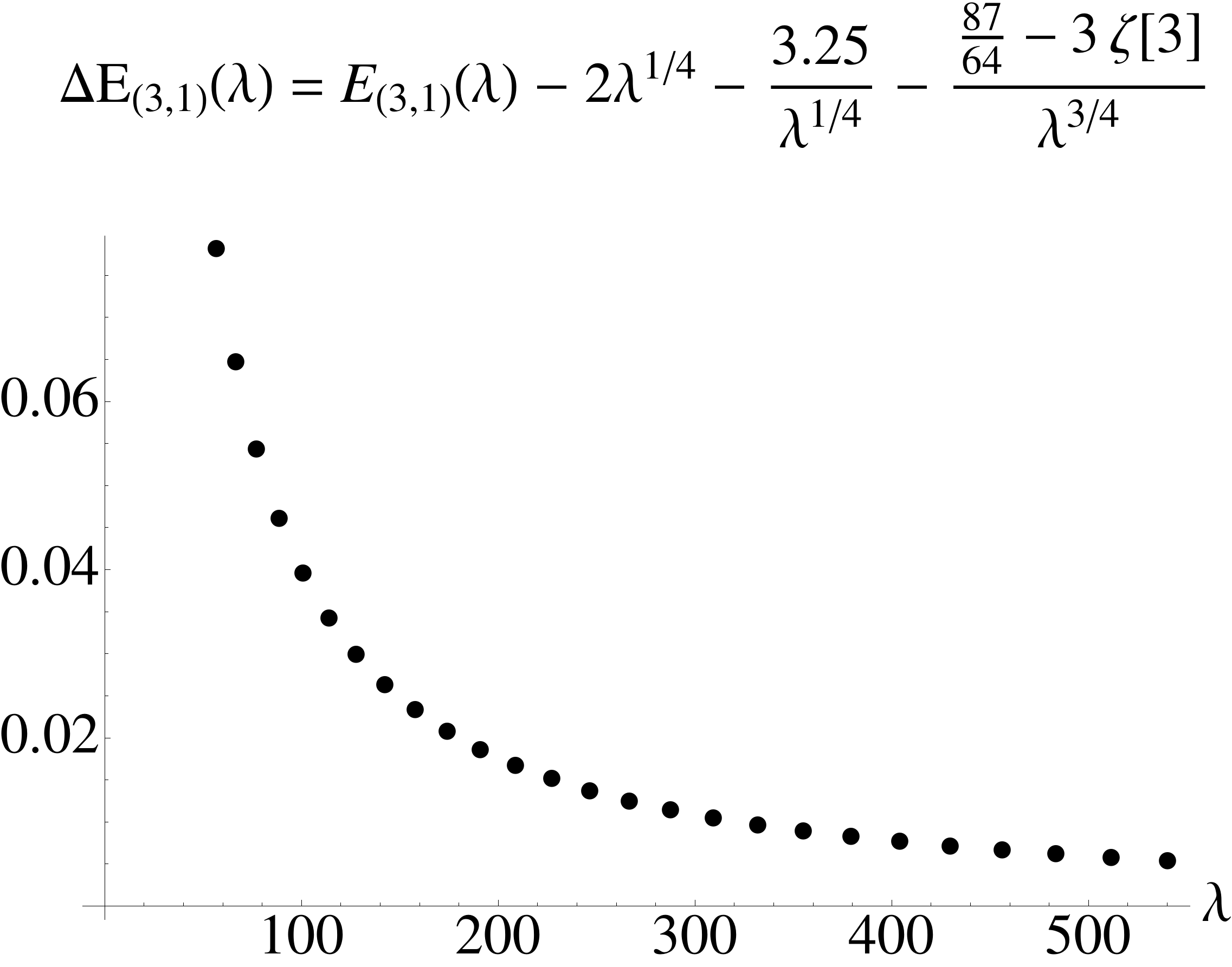}
\end{center}
\caption{\smaller 
Black dots represent the difference between the numerical solution and conjectured 
asymptotic expansions.
}
\la{figJ3n1}
\end{figure}The outcome of this fitting is very interesting because the coefficient $c_4$ becomes very small and this implies that up to the overall $\sqrt[4]\lam$  the expansion may indeed be in powers of $1/\sqrt\lam$.\footnote{ One can check by using the data from \cite{F10} that a similar phenomenon also happens for the Konishi operator.
For example fitting the data in the interval $g\in [1.8,5.]$ with $c_{-1}=c_1=2,c_0=c_2=0,c_3=\frac{1}{2}-3 \zeta (3)\approx 3.10617$, one gets
$$
E_K=2 \sqrt[4]{\lambda }+\frac{2}{\sqrt[4]{\lambda }}-\frac{3.10617}{\lambda ^{3/4}}+\frac{0.0175158}{\lambda }+\frac{19.2403}{\lambda ^{5/4}}-\frac{39.3787}{\lambda
   ^{3/2}}+\frac{36.5992}{\lambda ^{7/4}}\,,
$$
which gives additional evidence in favor of the $1/\sqrt\lam$ expansion at least for the $n=1$ operators. } Finally setting $c_{2k}=0$ one gets the following fitting for the $J=3,n=1$ state
{\smaller 
\bea\la{FitEJ3n1e}
\begin{array}{|c|c|c|}
\hline
g_0&\lam_0& {\rm Fit} \\\hline
1.2 & 56.8489 & 1.99982 \sqrt[4]{\lambda }+\frac{3.25805}{\sqrt[4]{\lambda }}-\frac{2.30759}{\lambda
   ^{3/4}}+\frac{13.7801}{\lambda ^{5/4}}-\frac{11.3876}{\lambda ^{7/4}} \\
 1.3 & 66.7185 & 1.9999 \sqrt[4]{\lambda }+\frac{3.25325}{\sqrt[4]{\lambda }}-\frac{2.20648}{\lambda
   ^{3/4}}+\frac{12.8714}{\lambda ^{5/4}}-\frac{8.44329}{\lambda ^{7/4}} \\
 1.4 & 77.3777 & 1.99988 \sqrt[4]{\lambda }+\frac{3.25475}{\sqrt[4]{\lambda }}-\frac{2.23931}{\lambda
   ^{3/4}}+\frac{13.1794}{\lambda ^{5/4}}-\frac{9.49031}{\lambda ^{7/4}} \\
 1.5 & 88.8264 & 1.9997 \sqrt[4]{\lambda }+\frac{3.2658}{\sqrt[4]{\lambda }}-\frac{2.48916}{\lambda
   ^{3/4}}+\frac{15.6166}{\lambda ^{5/4}}-\frac{18.1459}{\lambda ^{7/4}}
    \\\hline
   \end{array}~~~~
\eea
}In Figure \ref{figJ3n1} we plot the difference between the numerical solution and its large $\lam$ asymptotics 
$2\lam^{1/4}+{3.25\ov\lam^{1/4}}$ and $2\lam^{1/4}+{3.25\ov\lam^{1/4}}-{2.2468\ov\lam^{3/4}}$.

\subsection{$J=4,n=1$ operator}

In the table \eqref{E41data} we present the data for the $J=4,n=1$ state.
For this operator the data is known only up to $g=3.4$ with a gap between $g=1.63$ and $g=2.2$. The operator has a critical point at $g\approx 1.83$ and a subcritical one at $g\approx 2.19$ and our iterations did not converge for $g\in[1.64,2.1]$.
By this reason we fit the data  in the interval $g\in [g_0, 3.4]$  with the assumption that $c_{2k}=0$
{\smaller 
\bea\la{FitEJ4n1a}
\begin{array}{|c|c|c|}
\hline
g_0&\lam_0& {\rm Fit} \\\hline
 0.9 & 31.9775 & 2.00114 \sqrt[4]{\lambda }+\frac{4.96572}{\sqrt[4]{\lambda }}-\frac{2.24203}{\lambda
   ^{3/4}}+\frac{21.2561}{\lambda ^{5/4}}-\frac{52.8464}{\lambda ^{7/4}} \\
 1. & 39.4784 & 2.00071 \sqrt[4]{\lambda }+\frac{4.98755}{\sqrt[4]{\lambda }}-\frac{2.63256}{\lambda
   ^{3/4}}+\frac{24.1832}{\lambda ^{5/4}}-\frac{60.6194}{\lambda ^{7/4}} \\
 1.1 & 47.7689 & 1.9995 \sqrt[4]{\lambda }+\frac{5.05089}{\sqrt[4]{\lambda }}-\frac{3.81607}{\lambda
   ^{3/4}}+\frac{33.5122}{\lambda ^{5/4}}-\frac{86.8559}{\lambda ^{7/4}} \\
 1.2 & 56.8489 & 1.99575 \sqrt[4]{\lambda }+\frac{5.25258}{\sqrt[4]{\lambda }}-\frac{7.71075}{\lambda
   ^{3/4}}+\frac{65.4442}{\lambda ^{5/4}}-\frac{180.907}{\lambda ^{7/4}}
    \\\hline
   \end{array}~~~~
\eea
}We see that $c_{-1}$ is indeed close to $2$, and $c_1\approx 5$ as expected for the $J=4,n=1$ state. Let us then fix $c_{-1}=2, c_1= 5$ 
\begin{figure}[t]
\begin{center}
\includegraphics*[width=0.5\textwidth]{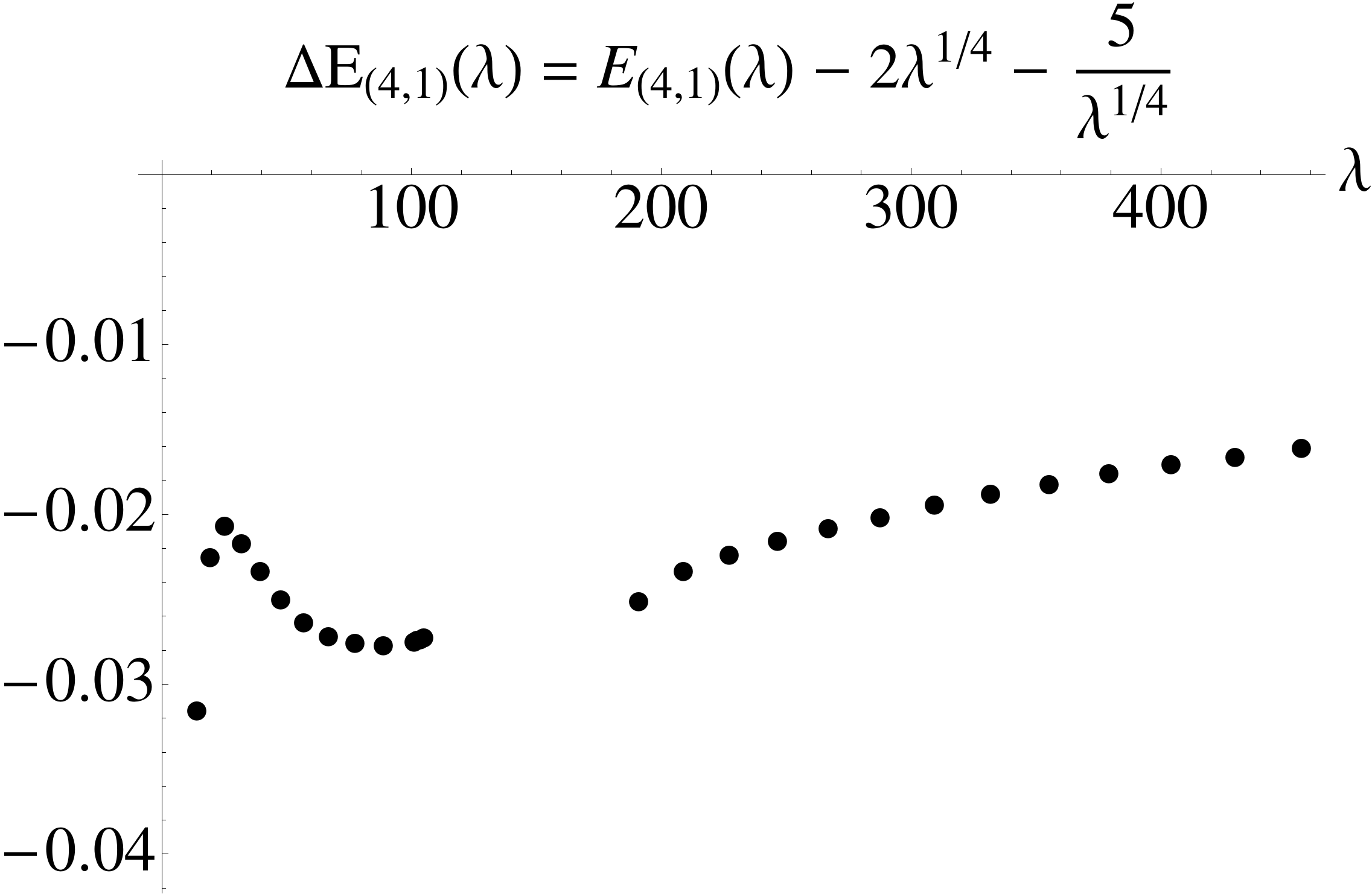}\qquad \includegraphics*[width=0.42\textwidth]{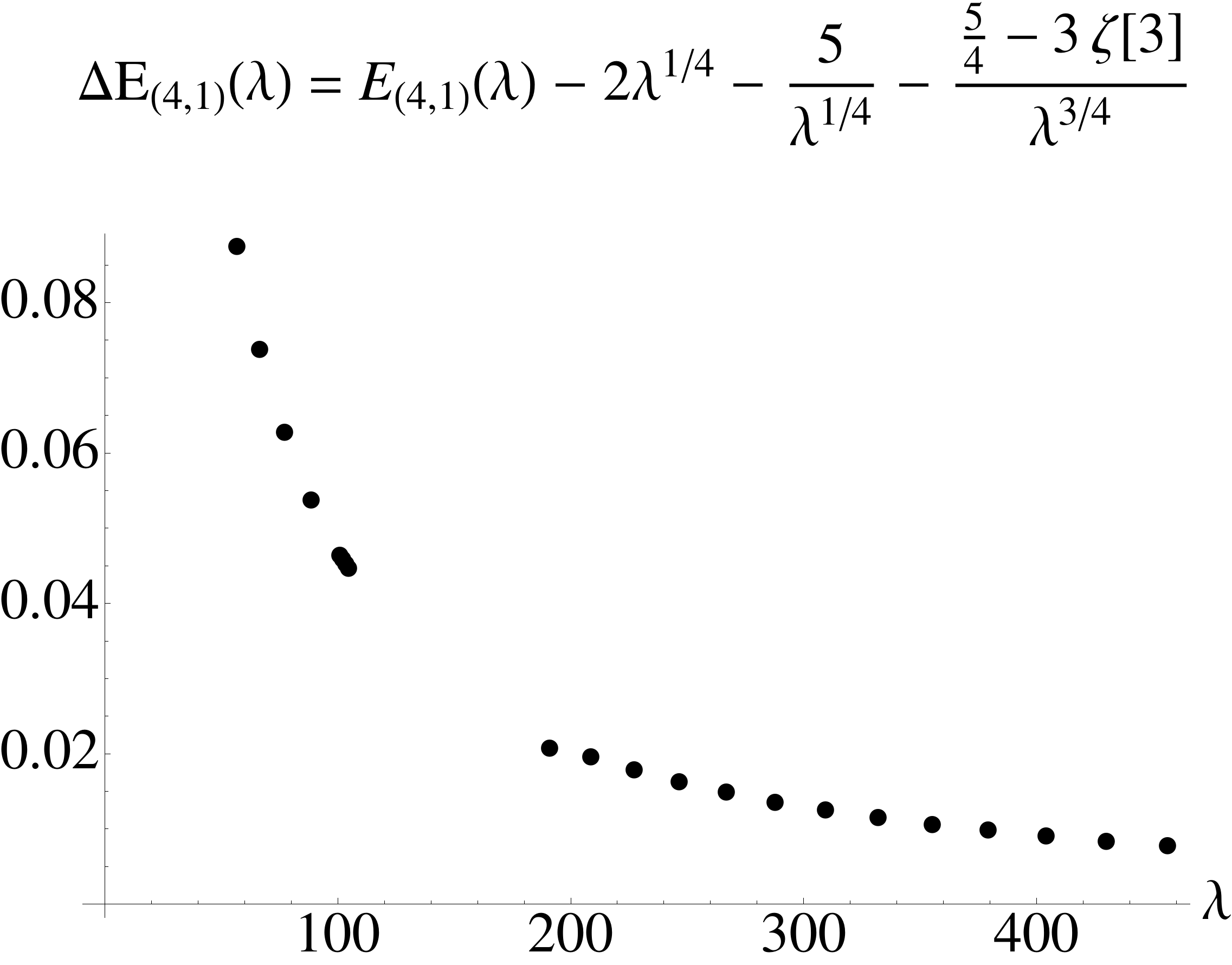}
\end{center}
\caption{\smaller 
Black dots represent the difference between the numerical solution and conjectured 
asymptotic expansions.
}
\la{figJ4n1}
\end{figure}
{\smaller 
\bea\la{FitEJ4n1b}
\begin{array}{|c|c|c|}
\hline
g_0&\lam_0& {\rm Fit} \\\hline
 0.9 & 31.9775 & 2 \sqrt[4]{\lambda }+\frac{5}{\sqrt[4]{\lambda }}-\frac{2.51332}{\lambda
   ^{3/4}}+\frac{21.4871}{\lambda ^{5/4}}-\frac{50.4063}{\lambda ^{7/4}} \\
 1. & 39.4784 & 2 \sqrt[4]{\lambda }+\frac{5}{\sqrt[4]{\lambda }}-\frac{2.49341}{\lambda
   ^{3/4}}+\frac{21.0556}{\lambda ^{5/4}}-\frac{48.2863}{\lambda ^{7/4}} \\
 1.1 & 47.7689 & 2 \sqrt[4]{\lambda }+\frac{5}{\sqrt[4]{\lambda }}-\frac{2.47161}{\lambda
   ^{3/4}}+\frac{20.5437}{\lambda ^{5/4}}-\frac{45.5455}{\lambda ^{7/4}} \\
 1.2 & 56.8489 & 2 \sqrt[4]{\lambda }+\frac{5}{\sqrt[4]{\lambda }}-\frac{2.44817}{\lambda
   ^{3/4}}+\frac{19.9559}{\lambda ^{5/4}}-\frac{42.1681}{\lambda ^{7/4}}
    \\\hline
   \end{array}~~~~
\eea
}According to \cite{Gr11} the coefficient $c_3$ for the $J=4,n=1$ state should be equal to  $c_3=\frac{5}{4}-3 \zeta (3)\approx -2.35617$, and it indeed agrees well with  the one we get from the fit. 
In Figure \ref{figJ4n1} we plot the difference between the numerical solution and its large $\lam$ asymptotics 
$2\lam^{1/4}+{5\ov\lam^{1/4}}$ and $2\lam^{1/4}+{5\ov\lam^{1/4}}-{2.35617\ov\lam^{3/4}}$. 

\begin{figure}[t]
\begin{center}
\includegraphics*[width=0.49\textwidth]{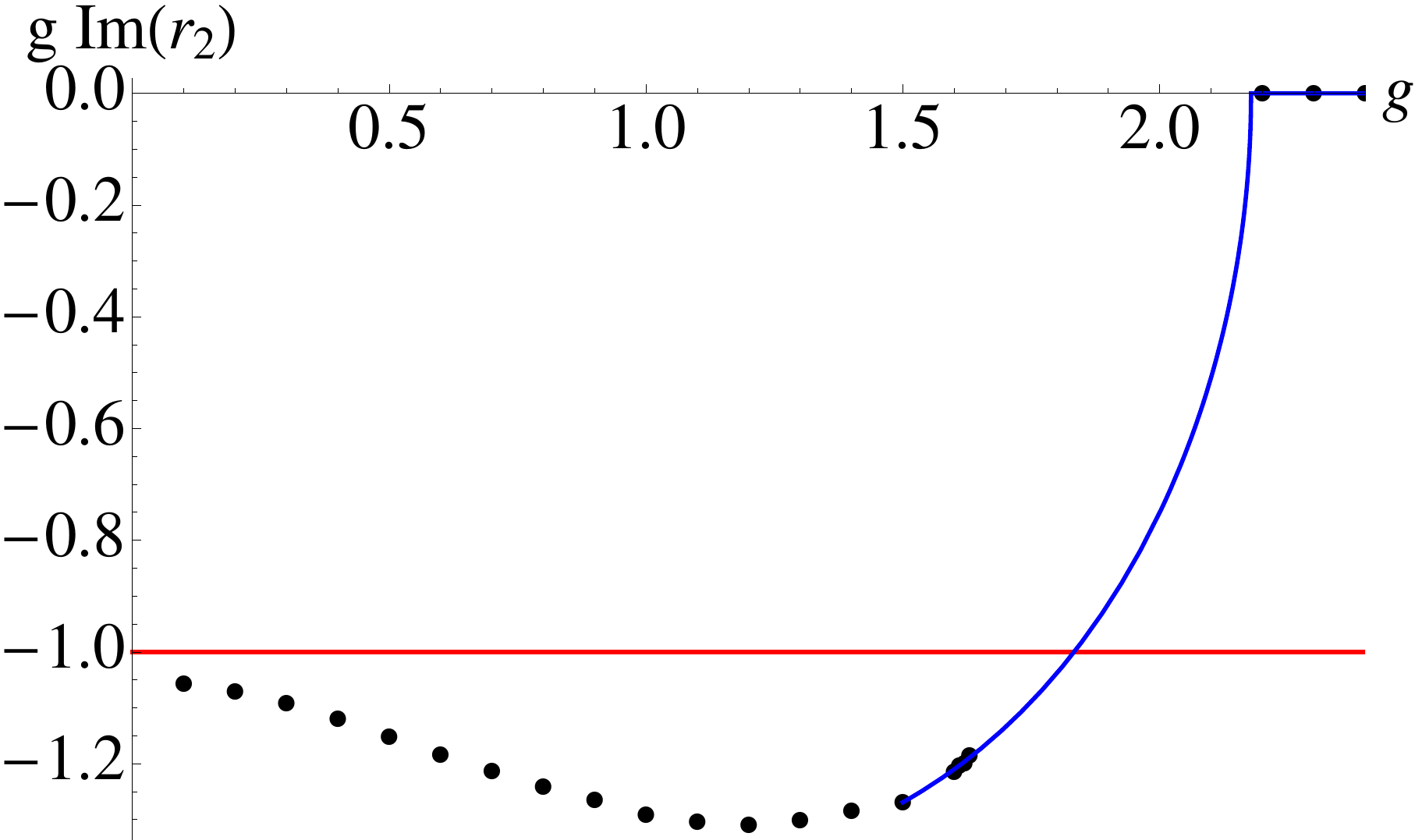}\qquad \includegraphics*[width=0.45\textwidth]{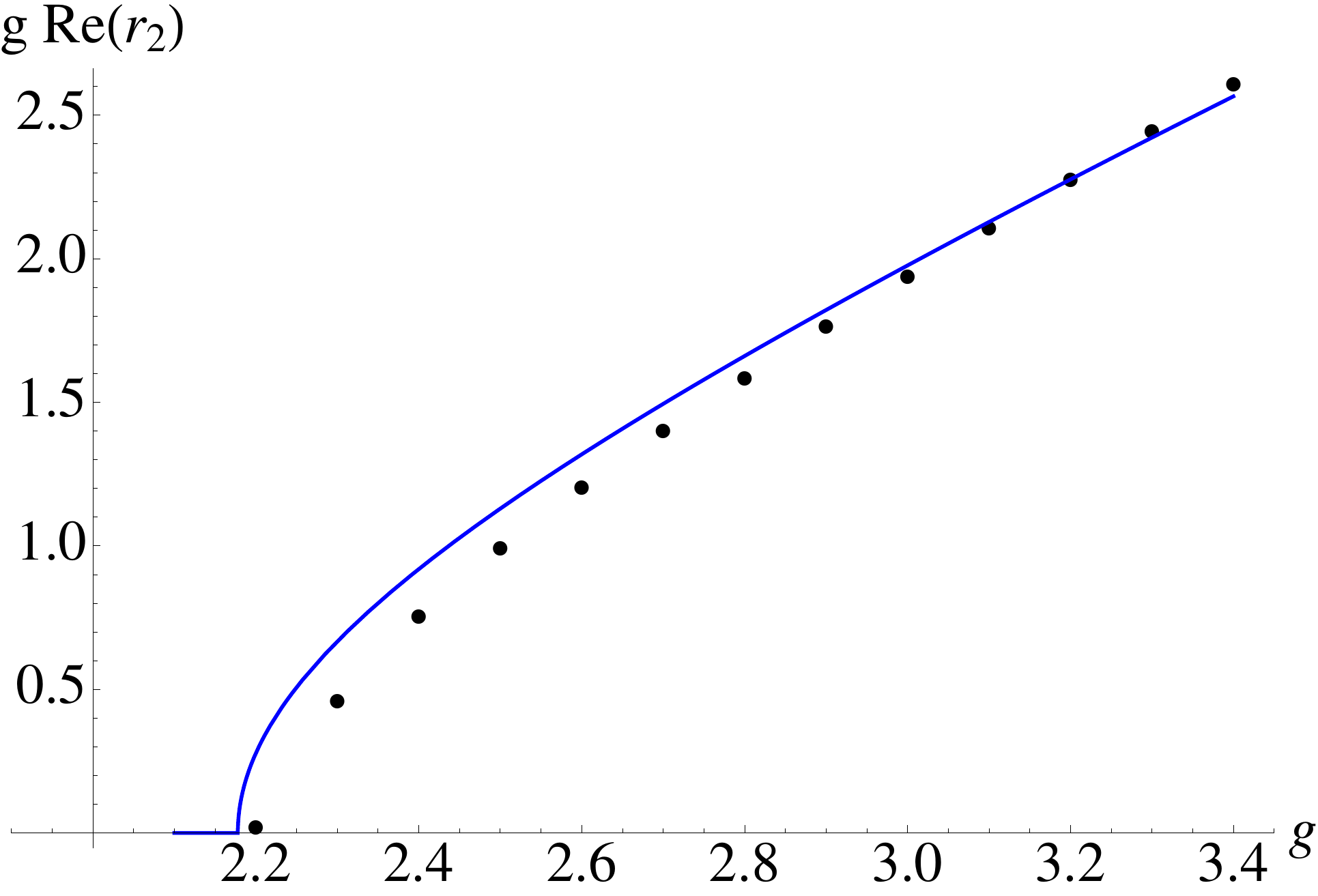}
\end{center}
\caption{\smaller  On the left and right figures the graphs of the imaginary and real parts of the rescaled root $r_2$ together with the fitting function (blue curves) $-1.25844\, i\, \sqrt{g(2.17804-g)}$ are shown.
}
\la{r2graph}
\end{figure}

To determine the values of the critical $g_{cr}$ and subcritical $\bar g_{cr}$ points we found the solutions to the equation $Y_{1|vw}(r_2 +{i\ov g})=-1$, where $r_2$ is also a root of $Y_{2|vw}$. The results are collected in Table \ref{r2data} and shown in Figure \ref{r2graph}.
{\smaller \bea\la{r2data}
\begin{array}{|c|c||}
\hline
g& r_2
\\
\hline
 0.1 & -10.562 i \\
 0.2 & -5.34994 i \\
 0.3 & -3.63868 i \\
 0.4 & -2.7981 i \\
 0.5 & -2.30147 i \\
 0.6 & -1.97299 i \\
 0.7 & -1.73166 i 
 \\\hline
\end{array}\begin{array}{|c|c||}
\hline
g& r_2
\\
\hline
 0.8 & -1.55116 i \\
 0.9 & -1.40522 i \\
 1. & -1.29021 i \\
 1.1 & -1.18521 i \\
 1.2 & -1.09082 i \\
 1.3 & -1.00062 i \\
 1.4 & -0.91655 i 
 \\\hline
\end{array}
\begin{array}{|c|c||}
\hline
g& r_2
\\
\hline
 1.5 & -0.84559 i\\
 1.6 & -0.75899 i \\
 1.61 & -0.74687 i \\
 1.62 & -0.739759 i \\
 1.63 & -0.727045 i \\
&\\
&
  \\\hline
\end{array}
\begin{array}{|c|c||}
\hline
g& r_2
\\
\hline
 2.2 & 0.0087264 \\
 2.3 & 0.199508 \\
 2.4 & 0.314217 \\
 2.5 & 0.397187 \\
 2.6 & 0.463024 \\
 2.7 & 0.519 \\
 2.8 & 0.565948 
 \\\hline
\end{array}
\begin{array}{|c|c||}
\hline
g& r_2
\\
\hline
 2.9 & 0.608164 \\
 3. & 0.646168 \\
 3.1 & 0.679729 \\
 3.2 & 0.711265 \\
 3.3 & 0.740638 \\
 3.4 & 0.767281\\
 &
 \\\hline
\end{array}
\eea
}
The root $r_2$ is purely imaginary for $g\le 2.1$ and real for $g\ge 2.2$. It vanishes at $g=\bar{g}_{cr}$ 
Fitting our data  in the interval $[1.5,16.3]$ to the function $c\sqrt{g-\bar{g}_{cr}\ov g}$, we get
\be
r_2(g)\sim -1.25844\, i\, \sqrt{2.17804-g\ov g}\,.
\ee
The subcritical value obtained from this fitting agrees very well with the data in Table \ref{r2data} which shows that it should be $\bar{g}_{cr}\approx 2.19$.  
To find the critical value we use the fitting function and solve the equation $r_2(g)=-i/g$. This gives ${g}_{cr}\approx 1.83$.

\subsection{$J=4,n=2$ operator}

In the table \eqref{E42data} we present the results of the computation of the energy of the $J=4,n=2$ state.
Fitting the data in the interval $g\in [g_0, 7.7]$ we get 
{\smaller 
\bea\la{FitEJ4n2a}
\begin{array}{|c|c|c|}
\hline
g_0&\lam_0& {\rm Fit} \\\hline
 1.6 & 101.065 & 2.01324 \sqrt[4]{\Lambda }-0.599655+\frac{16.0334}{\sqrt[4]{\Lambda }}-\frac{106.352}{\sqrt{\Lambda
   }}+\frac{535.582}{\Lambda ^{3/4}}-\frac{1548.42}{\Lambda }+\frac{1884.37}{\Lambda ^{5/4}} \\
 1.7 & 114.093 & 2.00765 \sqrt[4]{\Lambda }-0.364641+\frac{11.9675}{\sqrt[4]{\Lambda }}-\frac{69.3126}{\sqrt{\Lambda
   }}+\frac{348.21}{\Lambda ^{3/4}}-\frac{1049.34}{\Lambda }+\frac{1337.51}{\Lambda ^{5/4}} \\
 1.8 & 127.91 & 2.00248 \sqrt[4]{\Lambda }-0.144622+\frac{8.11028}{\sqrt[4]{\Lambda }}-\frac{33.6705}{\sqrt{\Lambda
   }}+\frac{165.145}{\Lambda ^{3/4}}-\frac{553.8}{\Lambda }+\frac{785.161}{\Lambda ^{5/4}} \\
 1.9 & 142.517 & 1.99906 \sqrt[4]{\Lambda }+0.00296822+\frac{5.49016}{\sqrt[4]{\Lambda }}-\frac{9.13276}{\sqrt{\Lambda
   }}+\frac{37.2992}{\Lambda ^{3/4}}-\frac{202.443}{\Lambda }+\frac{387.196}{\Lambda ^{5/4}}    \\\hline
   \end{array}~~~~
\eea
}where the expansion parameter $\Lambda = n^2\lambda = 4\lambda$ is introduced.
One sees that $c_{-1}$ is very close to 2 as expected, and 
fixing  $c_{-1}= 2$ one gets
{\smaller 
\bea\la{FitEJ4n2b}
\begin{array}{|c|c|c|}
\hline
g_0&\lam_0& {\rm Fit} \\\hline
1.6 & 101.065 & 2 \sqrt[4]{\Lambda }-0.0636918+\frac{7.12558}{\sqrt[4]{\Lambda }}-\frac{28.5421}{\sqrt{\Lambda
   }}+\frac{158.802}{\Lambda ^{3/4}}-\frac{589.219}{\Lambda }+\frac{881.06}{\Lambda ^{5/4}} \\
 1.7 & 114.093 & 2 \sqrt[4]{\Lambda }-0.0505662+\frac{6.66846}{\sqrt[4]{\Lambda }}-\frac{22.2713}{\sqrt{\Lambda
   }}+\frac{116.455}{\Lambda ^{3/4}}-\frac{448.43}{\Lambda }+\frac{696.68}{\Lambda ^{5/4}} \\
 1.8 & 127.91 & 2 \sqrt[4]{\Lambda }-0.0413324+\frac{6.34268}{\sqrt[4]{\Lambda }}-\frac{17.7387}{\sqrt{\Lambda
   }}+\frac{85.3727}{\Lambda ^{3/4}}-\frac{343.379}{\Lambda }+\frac{556.661}{\Lambda ^{5/4}} \\
 1.9 & 142.517 & 2 \sqrt[4]{\Lambda }-0.0367516+\frac{6.17907}{\sqrt[4]{\Lambda }}-\frac{15.4318}{\sqrt{\Lambda
   }}+\frac{69.3241}{\Lambda ^{3/4}}-\frac{288.293}{\Lambda }+\frac{482.019}{\Lambda ^{5/4}}
  \\\hline
   \end{array}~~~~
\eea
}We see that the coefficient $c_0$ becomes small and we set it to 0: $c_0=0$
{\smaller 
\bea\la{FitEJ4n2c}
\begin{array}{|c|c|c|}
\hline
g_0&\lam_0& {\rm Fit} \\\hline
  1.6 & 101.065 & 2 \sqrt[4]{\Lambda }+\frac{5.0046}{\sqrt[4]{\Lambda }}-\frac{0.795942}{\sqrt{\Lambda
   }}-\frac{19.4612}{\Lambda ^{3/4}}-\frac{26.4872}{\Lambda }+\frac{182.368}{\Lambda ^{5/4}} \\
 1.7 & 114.093 & 2 \sqrt[4]{\Lambda }+\frac{4.95929}{\sqrt[4]{\Lambda }}+\frac{0.454694}{\sqrt{\Lambda
   }}-\frac{32.1551}{\Lambda ^{3/4}}+\frac{29.665}{\Lambda }+\frac{90.9888}{\Lambda ^{5/4}} \\
 1.8 & 127.91 & 2 \sqrt[4]{\Lambda }+\frac{4.92584}{\sqrt[4]{\Lambda }}+\frac{1.39108}{\sqrt{\Lambda
   }}-\frac{41.8075}{\Lambda ^{3/4}}+\frac{73.0916}{\Lambda }+\frac{19.0113}{\Lambda ^{5/4}} \\
 1.9 & 142.517 & 2 \sqrt[4]{\Lambda }+\frac{4.90238}{\sqrt[4]{\Lambda }}+\frac{2.05656}{\sqrt{\Lambda
   }}-\frac{48.7676}{\Lambda ^{3/4}}+\frac{104.906}{\Lambda }-\frac{34.6303}{\Lambda ^{5/4}}
    \\\hline
   \end{array}~~~~
\eea
}This fitting shows that $c_1\approx 5$, and assuming that $b_1$ is independent of $J$ one concludes that  for any two-particle $n=2$ state we should expect the same formula as for the $n=1$ states: $c_1(J,2)=J^2/4 + 1$.
Setting $c_1 = 5$, one gets
{\smaller 
\bea\la{FitEJ4n2d}
\begin{array}{|c|c|c|}
\hline
g_0&\lam_0& {\rm Fit} \\\hline
 1.6 & 101.065 & 2 \sqrt[4]{\Lambda }+\frac{5}{\sqrt[4]{\Lambda }}-\frac{0.675854}{\sqrt{\Lambda
   }}-\frac{20.6105}{\Lambda ^{3/4}}-\frac{21.7064}{\Lambda }+\frac{175.066}{\Lambda ^{5/4}} \\
 1.7 & 114.093 & 2 \sqrt[4]{\Lambda }+\frac{5}{\sqrt[4]{\Lambda }}-\frac{0.626357}{\sqrt{\Lambda
   }}-\frac{21.6187}{\Lambda ^{3/4}}-\frac{15.0422}{\Lambda }+\frac{160.75}{\Lambda ^{5/4}} \\
 1.8 & 127.91 & 2 \sqrt[4]{\Lambda }+\frac{5}{\sqrt[4]{\Lambda }}-\frac{0.608926}{\sqrt{\Lambda
   }}-\frac{21.9793}{\Lambda ^{3/4}}-\frac{12.6155}{\Lambda }+\frac{155.435}{\Lambda ^{5/4}} \\
 1.9 & 142.517 & 2 \sqrt[4]{\Lambda }+\frac{5}{\sqrt[4]{\Lambda }}-\frac{0.614803}{\sqrt{\Lambda
   }}-\frac{21.856}{\Lambda ^{3/4}}-\frac{13.4596}{\Lambda }+\frac{157.318}{\Lambda ^{5/4}}
    \\\hline
   \end{array}~~~~
\eea
}The coefficient $c_2$ is not really small and we cannot reliably conclude that it vanishes.  The contribution of the corresponding term is however smaller than the contribution of the next term and we believe that increasing the precision of the computation one would show that  $c_2=0$. 

\begin{figure}[t]
\begin{center}
\includegraphics*[width=0.475\textwidth]{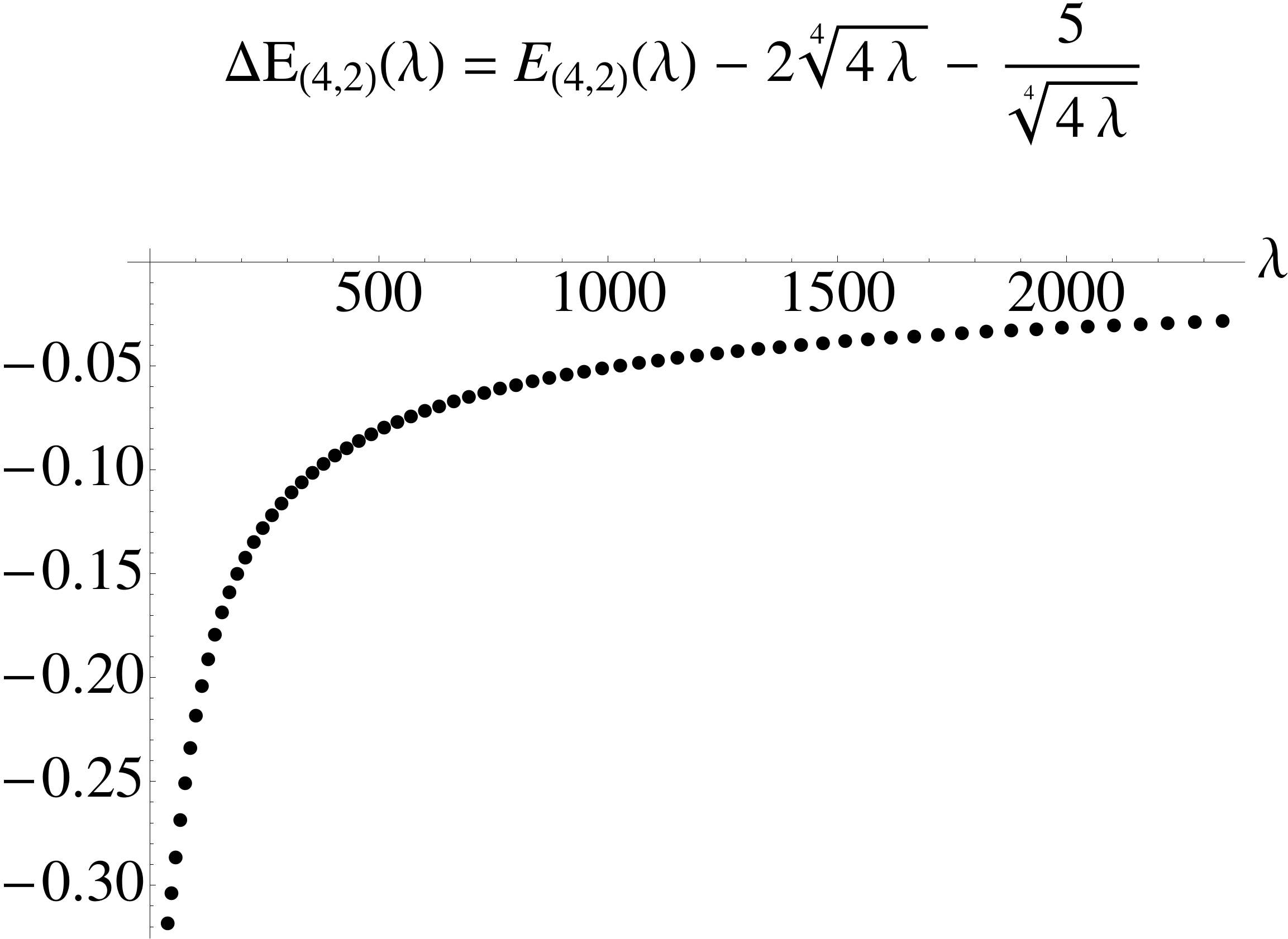}\qquad \includegraphics*[width=0.45\textwidth]{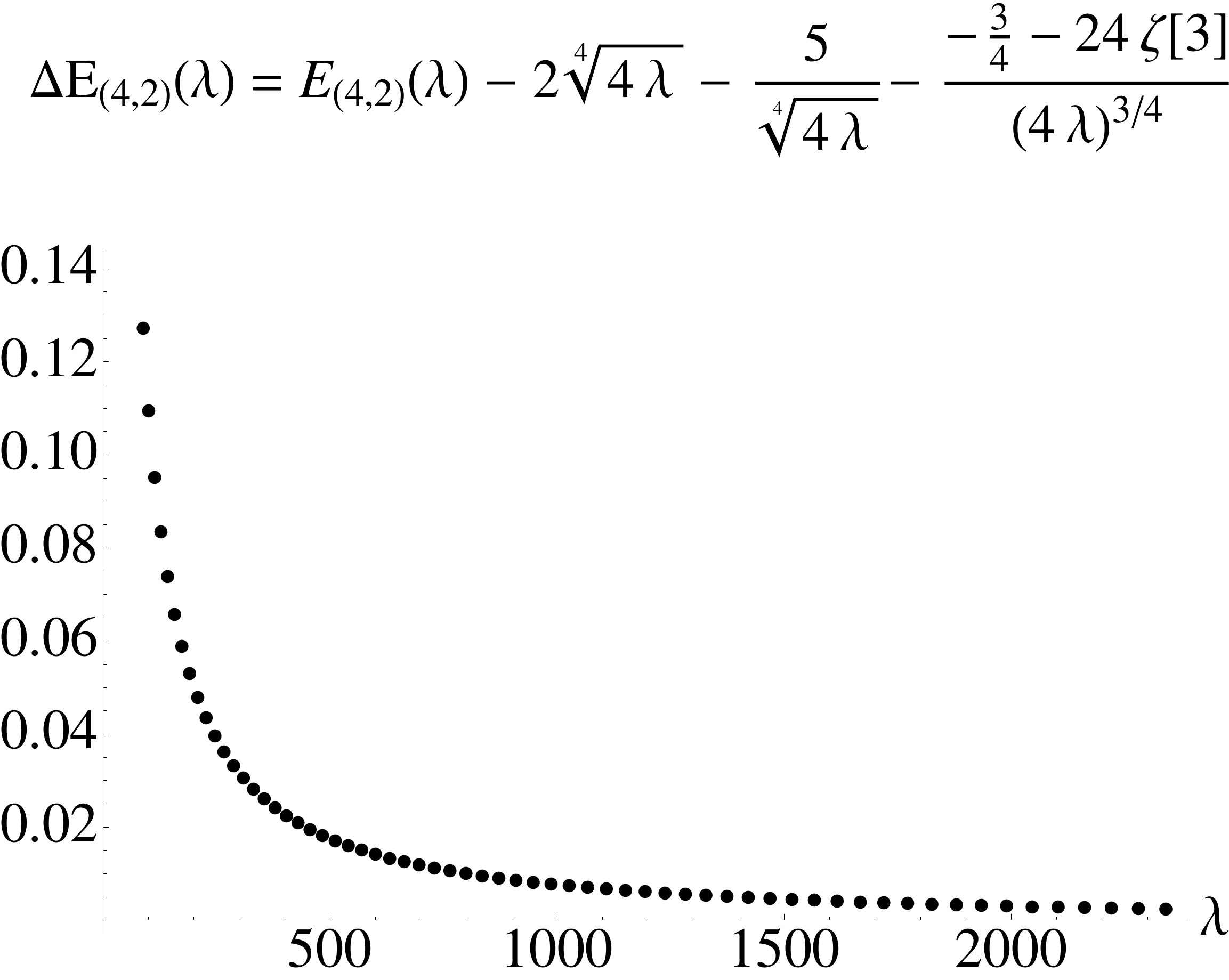}
\end{center}
\caption{\smaller 
Black dots represent the difference between the numerical solution and conjectured 
asymptotic expansions.
}
\la{figJ4n2}
\end{figure}

Assuming that $c_{2k}=0$ one gets the following fitting
{\smaller 
\bea\la{FitEJ4n2e}
\begin{array}{|c|c|c|}
\hline
g_0&\lam_0& {\rm Fit} \\\hline
 1.6 & 101.065 & 2.00005 \sqrt[4]{\Lambda }+\frac{4.96322}{\sqrt[4]{\Lambda }}-\frac{26.0062}{\Lambda
   ^{3/4}}+\frac{136.487}{\Lambda ^{5/4}}+\frac{65.3371}{\Lambda ^{7/4}} \\
 1.7 & 114.093 & 1.99999 \sqrt[4]{\Lambda }+\frac{4.97512}{\sqrt[4]{\Lambda }}-\frac{26.8411}{\Lambda
   ^{3/4}}+\frac{160.533}{\Lambda ^{5/4}}-\frac{175.699}{\Lambda ^{7/4}} \\
 1.8 & 127.91 & 1.99995 \sqrt[4]{\Lambda }+\frac{4.98407}{\sqrt[4]{\Lambda }}-\frac{27.4881}{\Lambda
   ^{3/4}}+\frac{179.837}{\Lambda ^{5/4}}-\frac{377.218}{\Lambda ^{7/4}} \\
 1.9 & 142.517 & 1.99992 \sqrt[4]{\Lambda }+\frac{4.99048}{\sqrt[4]{\Lambda }}-\frac{27.9635}{\Lambda
   ^{3/4}}+\frac{194.497}{\Lambda ^{5/4}}-\frac{536.145}{\Lambda ^{7/4}}
    \\\hline
   \end{array}\,,~~~~
\eea
}which also agrees very well with $c_1=5$. The coefficient $c_3$ for the states with $n=2$ is conjectured \cite{Gr11}
 to be equal to 
\be\la{c3n2}
c_3=-\frac{J^4}{64}+\frac{3 J^2}{8}-24 \zeta (3)-\frac{11}{4}\,,
\ee
and for $J=4$, $c_3\approx -29.5994$. We see that the number is indeed close to the one we get from the fit. The agreement with the conjectured value of $c_3$ becomes even more evident if one sets $c_{-1}=2, c_1=5$
{\smaller 
\bea\la{FitEJ4n2f}
\begin{array}{|c|c|c|}
\hline
g_0&\lam_0& {\rm Fit} \\\hline
 1.6 & 101.065 & 2 \sqrt[4]{\Lambda }+\frac{5}{\sqrt[4]{\Lambda
   }}-\frac{29.8997}{\Lambda ^{3/4}}+\frac{269.391}{\Lambda
   ^{5/4}}-\frac{1346.11}{\Lambda ^{7/4}} \\
 1.7 & 114.093 & 2 \sqrt[4]{\Lambda }+\frac{5}{\sqrt[4]{\Lambda
   }}-\frac{30.0467}{\Lambda ^{3/4}}+\frac{281.386}{\Lambda
   ^{5/4}}-\frac{1561.66}{\Lambda ^{7/4}} \\
 1.8 & 127.91 & 2 \sqrt[4]{\Lambda }+\frac{5}{\sqrt[4]{\Lambda
   }}-\frac{30.1745}{\Lambda ^{3/4}}+\frac{292.273}{\Lambda
   ^{5/4}}-\frac{1767.33}{\Lambda ^{7/4}} \\
 1.9 & 142.517 & 2 \sqrt[4]{\Lambda }+\frac{5}{\sqrt[4]{\Lambda
   }}-\frac{30.2888}{\Lambda ^{3/4}}+\frac{302.41}{\Lambda
   ^{5/4}}-\frac{1967.94}{\Lambda ^{7/4}}    \\\hline
   \end{array}~~~~
\eea
}In Figure \ref{figJ4n2} we plot the difference between the numerical solution and its large $\lam$ asymptotics 
$2(4\lam)^{1/4}+{5\ov(4\lam)^{1/4}}$ and $2(4\lam)^{1/4}+{5\ov(4\lam)^{1/4}}-{29.6\ov (4\lam)^{3/4}}$.

\subsection{$J=5,n=2$ operator}

In the table \eqref{E52data} we present the results of the computation of the energy of the $J=5,n=2$ state.
Fitting the data in the interval $g\in [g_0, 7.7]$ we get 
{\smaller 
\bea\la{FitEJ5n2a}
\begin{array}{|c|c|c|}
\hline
g_0&\lam_0& {\rm Fit} \\\hline
1.6 & 101.065 & 2.00445 \sqrt[4]{\Lambda }-0.20912+\frac{11.1966}{\sqrt[4]{\Lambda }}-\frac{38.8728}{\sqrt{\Lambda
   }}+\frac{178.449}{\Lambda ^{3/4}}-\frac{577.756}{\Lambda }+\frac{881.1}{\Lambda ^{5/4}} \\
 1.7 & 114.093 & 2.00546 \sqrt[4]{\Lambda }-0.251818+\frac{11.9353}{\sqrt[4]{\Lambda }}-\frac{45.6022}{\sqrt{\Lambda
   }}+\frac{212.492}{\Lambda ^{3/4}}-\frac{668.43}{\Lambda }+\frac{980.454}{\Lambda ^{5/4}} \\
 1.8 & 127.91 & 2.00297 \sqrt[4]{\Lambda }-0.145733+\frac{10.0755}{\sqrt[4]{\Lambda }}-\frac{28.4169}{\sqrt{\Lambda
   }}+\frac{124.224}{\Lambda ^{3/4}}-\frac{429.497}{\Lambda }+\frac{714.13}{\Lambda ^{5/4}} \\
 1.9 & 142.517 & 1.99987 \sqrt[4]{\Lambda }-0.0122334+\frac{7.70547}{\sqrt[4]{\Lambda }}-\frac{6.22177}{\sqrt{\Lambda
   }}+\frac{8.58421}{\Lambda ^{3/4}}-\frac{111.684}{\Lambda }+\frac{354.159}{\Lambda ^{5/4}}
   \\\hline
   \end{array}~~~~
\eea
}where $\Lambda = n^2\lambda = 4\lambda$ is the same expansion parameter as for the $J=4,n=2$ state.
One sees that $c_{-1}$ is very close to 2 as expected, and 
fixing  $c_{-1}= 2$ one gets
{\smaller 
\bea\la{FitEJ5n2b}
\begin{array}{|c|c|c|}
\hline
g_0&\lam_0& {\rm Fit} \\\hline
 1.6 & 101.065 & 2 \sqrt[4]{\Lambda }-0.0290362+\frac{8.20356}{\sqrt[4]{\Lambda }}-\frac{12.7288}{\sqrt{\Lambda
   }}+\frac{51.8515}{\Lambda ^{3/4}}-\frac{255.467}{\Lambda }+\frac{543.99}{\Lambda ^{5/4}} \\
 1.7 & 114.093 & 2 \sqrt[4]{\Lambda }-0.0275291+\frac{8.15107}{\sqrt[4]{\Lambda }}-\frac{12.0088}{\sqrt{\Lambda
   }}+\frac{46.9889}{\Lambda ^{3/4}}-\frac{239.301}{\Lambda }+\frac{522.818}{\Lambda ^{5/4}} \\
 1.8 & 127.91 & 2 \sqrt[4]{\Lambda }-0.0221103+\frac{7.95989}{\sqrt[4]{\Lambda }}-\frac{9.34883}{\sqrt{\Lambda
   }}+\frac{28.7486}{\Lambda ^{3/4}}-\frac{177.652}{\Lambda }+\frac{440.648}{\Lambda ^{5/4}} \\
 1.9 & 142.517 & 2 \sqrt[4]{\Lambda }-0.0175866+\frac{7.79832}{\sqrt[4]{\Lambda }}-\frac{7.07073}{\sqrt{\Lambda
   }}+\frac{12.9004}{\Lambda ^{3/4}}-\frac{123.254}{\Lambda }+\frac{366.939}{\Lambda ^{5/4}}
     \\\hline
   \end{array}~~~~
\eea
}We see that the coefficient $c_0$ is even smaller than it was for  the $J=4,n=2$ state, and setting it to 0 one gets
{\smaller 
\bea\la{FitEJ5n2c}
\begin{array}{|c|c|c|}
\hline
g_0&\lam_0& {\rm Fit} \\\hline
1.6 & 101.065 & 2 \sqrt[4]{\Lambda }+\frac{7.23663}{\sqrt[4]{\Lambda }}-\frac{0.0797212}{\sqrt{\Lambda
   }}-\frac{29.4165}{\Lambda ^{3/4}}+\frac{1.07508}{\Lambda }+\frac{225.465}{\Lambda ^{5/4}} \\
 1.7 & 114.093 & 2 \sqrt[4]{\Lambda }+\frac{7.22057}{\sqrt[4]{\Lambda }}+\frac{0.363655}{\sqrt{\Lambda
   }}-\frac{33.9167}{\Lambda ^{3/4}}+\frac{20.9822}{\Lambda }+\frac{193.07}{\Lambda ^{5/4}} \\
 1.8 & 127.91 & 2 \sqrt[4]{\Lambda }+\frac{7.20197}{\sqrt[4]{\Lambda }}+\frac{0.884427}{\sqrt{\Lambda
   }}-\frac{39.2849}{\Lambda ^{3/4}}+\frac{45.134}{\Lambda }+\frac{153.039}{\Lambda ^{5/4}} \\
 1.9 & 142.517 & 2 \sqrt[4]{\Lambda }+\frac{7.18739}{\sqrt[4]{\Lambda }}+\frac{1.29794}{\sqrt{\Lambda
   }}-\frac{43.6097}{\Lambda ^{3/4}}+\frac{64.9025}{\Lambda }+\frac{119.708}{\Lambda ^{5/4}}
    \\\hline
   \end{array}~~~~
\eea
}This fitting shows that $c_1$ agrees with $c_1(J,2)=J^2/4 + 1$.
Setting $c_1 = 7.25$, one gets
{\smaller 
\bea\la{FitEJ5n2d}
\begin{array}{|c|c|c|}
\hline
g_0&\lam_0& {\rm Fit} \\\hline
1.6 & 101.065 & 2 \sqrt[4]{\Lambda }+\frac{7.25}{\sqrt[4]{\Lambda }}-\frac{0.428891}{\sqrt{\Lambda
   }}-\frac{26.0747}{\Lambda ^{3/4}}-\frac{12.8258}{\Lambda }+\frac{246.699}{\Lambda ^{5/4}} \\
 1.7 & 114.093 & 2 \sqrt[4]{\Lambda }+\frac{7.25}{\sqrt[4]{\Lambda }}-\frac{0.417828}{\sqrt{\Lambda }}-\frac{26.3}{\Lambda
   ^{3/4}}-\frac{11.3363}{\Lambda }+\frac{243.5}{\Lambda ^{5/4}} \\
 1.8 & 127.91 & 2 \sqrt[4]{\Lambda }+\frac{7.25}{\sqrt[4]{\Lambda }}-\frac{0.410937}{\sqrt{\Lambda
   }}-\frac{26.4426}{\Lambda ^{3/4}}-\frac{10.3768}{\Lambda }+\frac{241.398}{\Lambda ^{5/4}} \\
 1.9 & 142.517 & 2 \sqrt[4]{\Lambda }+\frac{7.25}{\sqrt[4]{\Lambda }}-\frac{0.415359}{\sqrt{\Lambda
   }}-\frac{26.3498}{\Lambda ^{3/4}}-\frac{11.0121}{\Lambda }+\frac{242.815}{\Lambda ^{5/4}}
  \\\hline
   \end{array}~~~~
\eea
}The coefficient $c_2$ is again smaller than the one for the $J=4,n=2$ case, and moreover the contribution of the corresponding term is much smaller than the contribution of the next one. It is possible that one cannot see that $c_2$ vanishes because of the exponentially suppressed corrections  at large $\lam$. These corrections decrease with $J$ increasing and this  would explain why for the $J=5,n=2$ case the coefficients $c_0$ and $c_2$ are closer to 0 than the ones for the $J=4,n=2$ state.

\begin{figure}[t]
\begin{center}
\includegraphics*[width=0.475\textwidth]{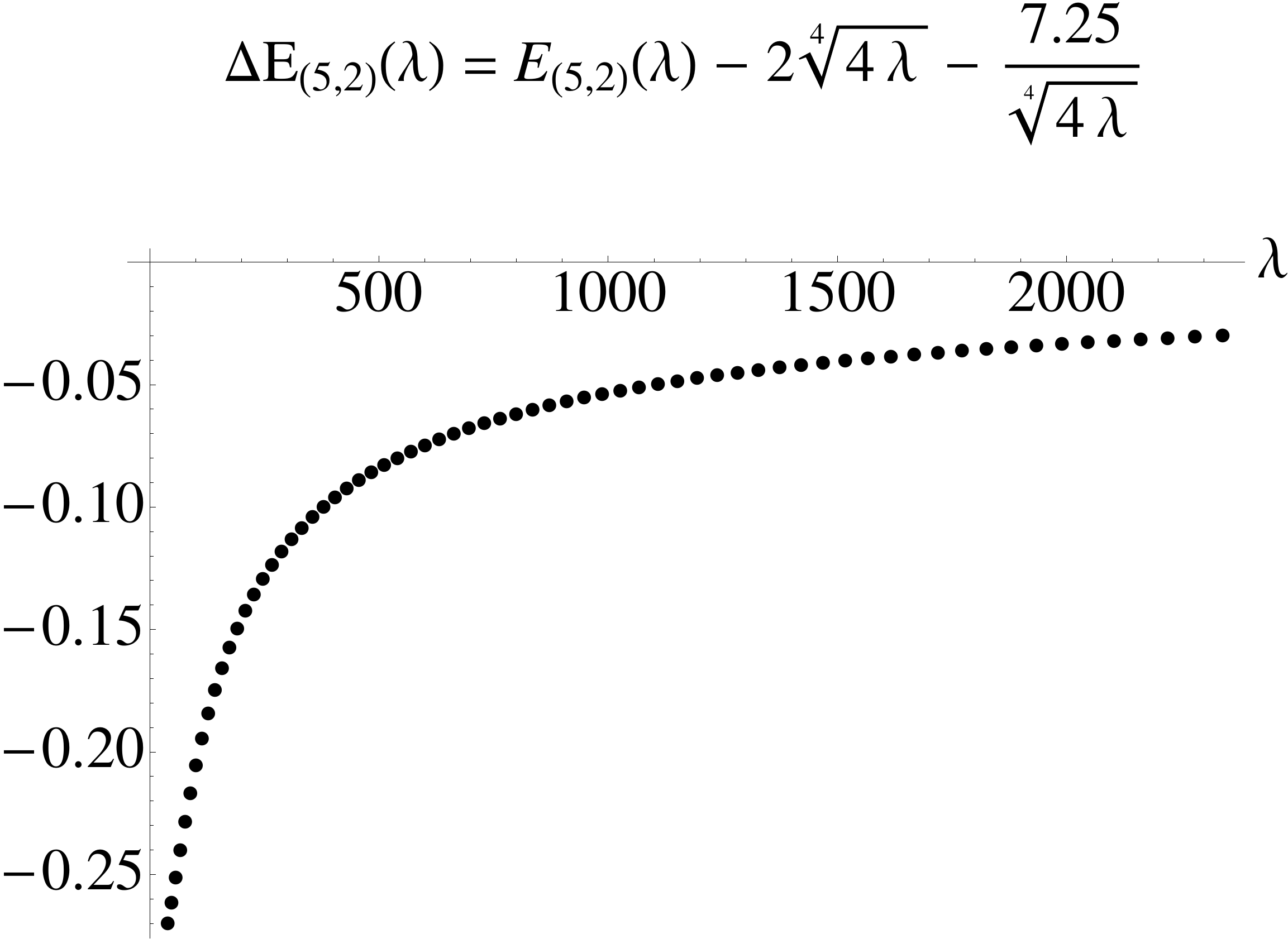}\qquad \includegraphics*[width=0.465\textwidth]{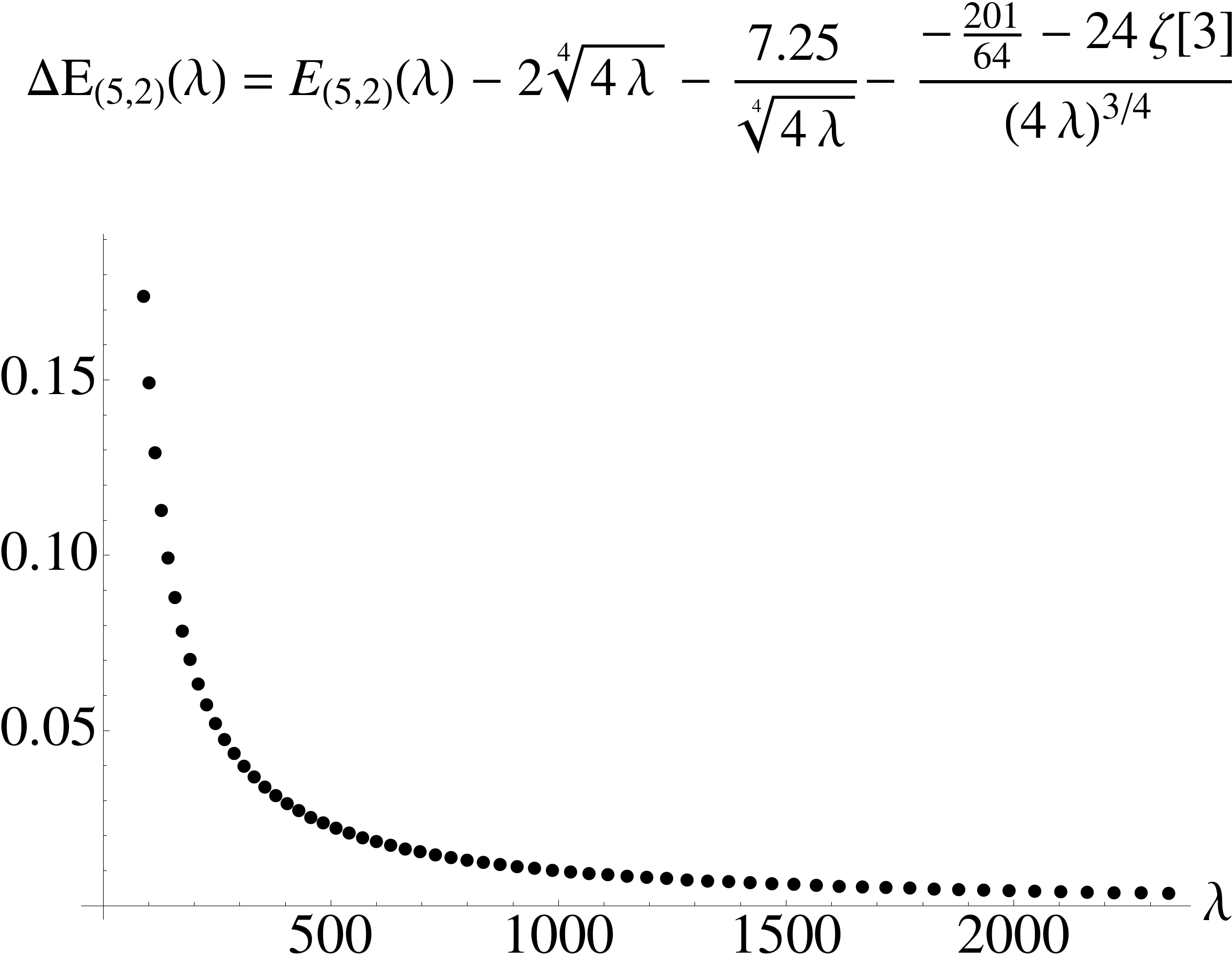}
\end{center}
\caption{\smaller 
Black dots represent the difference between the numerical solution and conjectured 
asymptotic expansions.
}
\la{figJ5n2}
\end{figure}

Next setting $c_{2k}=0$ one gets the following fitting
{\smaller 
\bea\la{FitEJ5n2e}
\begin{array}{|c|c|c|}
\hline
g_0&\lam_0& {\rm Fit} \\\hline
1.6 & 101.065 & 2.00001 \sqrt[4]{\Lambda }+\frac{7.23042}{\sqrt[4]{\Lambda }}-\frac{29.6628}{\Lambda
   ^{3/4}}+\frac{231.828}{\Lambda ^{5/4}}-\frac{27.3344}{\Lambda ^{7/4}} \\
 1.7 & 114.093 & 1.99999 \sqrt[4]{\Lambda }+\frac{7.23469}{\sqrt[4]{\Lambda }}-\frac{29.9624}{\Lambda
   ^{3/4}}+\frac{240.459}{\Lambda ^{5/4}}-\frac{113.852}{\Lambda ^{7/4}} \\
 1.8 & 127.91 & 1.99997 \sqrt[4]{\Lambda }+\frac{7.23966}{\sqrt[4]{\Lambda }}-\frac{30.3219}{\Lambda
   ^{3/4}}+\frac{251.185}{\Lambda ^{5/4}}-\frac{225.82}{\Lambda ^{7/4}} \\
 1.9 & 142.517 & 1.99995 \sqrt[4]{\Lambda }+\frac{7.24362}{\sqrt[4]{\Lambda }}-\frac{30.6156}{\Lambda
   ^{3/4}}+\frac{260.241}{\Lambda ^{5/4}}-\frac{323.995}{\Lambda ^{7/4}}
    \\\hline
   \end{array}\,,~~~~
\eea
}which agrees very well with $c_1=7.25$
and shows that the coefficient $c_3$ is close to the conjectured value
$c_3=-\frac{201}{64}-24 \zeta (3)\approx -31.99$.
Setting $c_{-1}=2, c_1=7.25$
{\smaller 
\bea\la{FitEJ5n2f}
\begin{array}{|c|c|c|}
\hline
g_0&\lam_0& {\rm Fit} \\\hline
  1.6 & 101.065 & 2 \sqrt[4]{\Lambda }+\frac{7.25}{\sqrt[4]{\Lambda }}-\frac{31.9121}{\Lambda
   ^{3/4}}+\frac{310.838}{\Lambda ^{5/4}}-\frac{878.22}{\Lambda ^{7/4}} \\
 1.7 & 114.093 & 2 \sqrt[4]{\Lambda }+\frac{7.25}{\sqrt[4]{\Lambda }}-\frac{32.}{\Lambda ^{3/4}}+\frac{318.014}{\Lambda
   ^{5/4}}-\frac{1007.16}{\Lambda ^{7/4}} \\
 1.8 & 127.91 & 2 \sqrt[4]{\Lambda }+\frac{7.25}{\sqrt[4]{\Lambda }}-\frac{32.0823}{\Lambda ^{3/4}}+\frac{325.019}{\Lambda
   ^{5/4}}-\frac{1139.5}{\Lambda ^{7/4}} \\
 1.9 & 142.517 & 2 \sqrt[4]{\Lambda }+\frac{7.25}{\sqrt[4]{\Lambda }}-\frac{32.1572}{\Lambda
   ^{3/4}}+\frac{331.669}{\Lambda ^{5/4}}-\frac{1271.1}{\Lambda ^{7/4}}
   \\\hline
   \end{array}~~~~
\eea
}makes the agreement even more impressive. 

In Figure \ref{figJ5n2} we plot the difference between the numerical solution and its large $\lam$ asymptotics 
 $2(4\lam)^{1/4}+{7.25\ov(4\lam)^{1/4}}$ and $2(4\lam)^{1/4}+{7.25\ov(4\lam)^{1/4}}-{31.99\ov(4\lam)^{3/4}}$ .

\subsection{$J=6,n=3$ operator}

In the table \eqref{E63data} we present the results of the computation of the energy of the $J=6,n=3$ state.
Fitting the data in the interval $g\in [g_0, 7.7]$ we get 
{\smaller 
\bea\la{FitEJ6n3a}
\begin{array}{|c|c|c|}
\hline
g_0&\lam_0& {\rm Fit} \\\hline
2.2 & 191.076 & 2.02122 \sqrt[4]{\Lambda }-1.1605+\frac{35.1776}{\sqrt[4]{\Lambda }}-\frac{312.4}{\sqrt{\Lambda
   }}+\frac{1974.72}{\Lambda ^{3/4}}-\frac{7098.66}{\Lambda }+\frac{10856.1}{\Lambda ^{5/4}} \\
 2.3 & 208.841 & 2.01305 \sqrt[4]{\Lambda }-0.711293+\frac{24.9684}{\sqrt[4]{\Lambda }}-\frac{189.655}{\sqrt{\Lambda
   }}+\frac{1151.37}{\Lambda ^{3/4}}-\frac{4177.2}{\Lambda }+\frac{6572.03}{\Lambda ^{5/4}} \\
 2.4 & 227.396 & 2.00352 \sqrt[4]{\Lambda }-0.182115+\frac{12.8223}{\sqrt[4]{\Lambda }}-\frac{42.0904}{\sqrt{\Lambda
   }}+\frac{150.596}{\Lambda ^{3/4}}-\frac{584.867}{\Lambda }+\frac{1239.93}{\Lambda ^{5/4}} \\
 2.5 & 246.74 & 1.99454 \sqrt[4]{\Lambda }+0.320466+\frac{1.17703}{\sqrt[4]{\Lambda }}+\frac{100.809}{\sqrt{\Lambda
   }}-\frac{828.784}{\Lambda ^{3/4}}+\frac{2969.68}{\Lambda }-\frac{4097.42}{\Lambda ^{5/4}} \\\hline
   \end{array}~~~~
\eea
}where the expansion parameter $\Lambda = n^2\lambda = 9\lambda$ is introduced.
One sees that  as expected $c_{-1}$ is close to 2, and 
fixing  $c_{-1}= 2$ one gets
{\smaller 
\bea\la{FitEJ6n3b}
\begin{array}{|c|c|c|}
\hline
g_0&\lam_0& {\rm Fit} \\\hline
2.2 & 191.076 & 2 \sqrt[4]{\Lambda }-0.0280675+\frac{10.2297}{\sqrt[4]{\Lambda }}-\frac{21.9871}{\sqrt{\Lambda
   }}+\frac{90.6627}{\Lambda ^{3/4}}-\frac{639.338}{\Lambda }+\frac{1711.89}{\Lambda ^{5/4}} \\
 2.3 & 208.841 & 2 \sqrt[4]{\Lambda }-0.00754532+\frac{9.29219}{\sqrt[4]{\Lambda }}-\frac{5.02319}{\sqrt{\Lambda
   }}-\frac{61.3087}{\Lambda ^{3/4}}+\frac{34.676}{\Lambda }+\frac{527.865}{\Lambda ^{5/4}} \\
 2.4 & 227.396 & 2 \sqrt[4]{\Lambda }+0.00938827+\frac{8.51111}{\sqrt[4]{\Lambda }}+\frac{9.25714}{\sqrt{\Lambda
   }}-\frac{190.658}{\Lambda ^{3/4}}+\frac{615.119}{\Lambda }-\frac{504.505}{\Lambda ^{5/4}} \\
 2.5 & 246.74 & 2 \sqrt[4]{\Lambda }+0.0202856+\frac{8.00379}{\sqrt[4]{\Lambda }}+\frac{18.6248}{\sqrt{\Lambda
   }}-\frac{276.408}{\Lambda ^{3/4}}+\frac{1004.23}{\Lambda }-\frac{1204.79}{\Lambda ^{5/4}}
  \\\hline
   \end{array}~~~~
\eea
}We see that the coefficient $c_0$ becomes small and we set it to 0: $c_0=0$
{\smaller 
\bea\la{FitEJ6n3c}
\begin{array}{|c|c|c|}
\hline
g_0&\lam_0& {\rm Fit} \\\hline
 2.2 & 191.076 & 2 \sqrt[4]{\Lambda }+\frac{8.99159}{\sqrt[4]{\Lambda }}-\frac{0.390111}{\sqrt{\Lambda
   }}-\frac{95.5746}{\Lambda ^{3/4}}+\frac{154.709}{\Lambda }+\frac{372.41}{\Lambda ^{5/4}} \\
 2.3 & 208.841 & 2 \sqrt[4]{\Lambda }+\frac{8.95568}{\sqrt[4]{\Lambda }}+\frac{0.91621}{\sqrt{\Lambda
   }}-\frac{113.173}{\Lambda ^{3/4}}+\frac{258.775}{\Lambda }+\frac{144.476}{\Lambda ^{5/4}} \\
 2.4 & 227.396 & 2 \sqrt[4]{\Lambda }+\frac{8.93423}{\sqrt[4]{\Lambda }}+\frac{1.70451}{\sqrt{\Lambda
   }}-\frac{123.911}{\Lambda ^{3/4}}+\frac{323.035}{\Lambda }+\frac{1.91857}{\Lambda ^{5/4}} \\
 2.5 & 246.74 & 2 \sqrt[4]{\Lambda }+\frac{8.9273}{\sqrt[4]{\Lambda }}+\frac{1.96172}{\sqrt{\Lambda
   }}-\frac{127.452}{\Lambda ^{3/4}}+\frac{344.466}{\Lambda }-\frac{46.2066}{\Lambda ^{5/4}}
    \\\hline
   \end{array}~~~~
\eea
}This fitting shows that $c_1\approx 9$, and assuming that $b_1$ is independent of $J$ one concludes that  for any two-particle $n=3$ state we should expect the  formula: $c_1(J,3)=J^2/4$. This is different from the $n=1$ and $n=2$ cases and therefore $c_1$ shows a nontrivial dependence of the string level $n$.
\begin{figure}[t]
\begin{center}
\includegraphics*[width=0.475\textwidth]{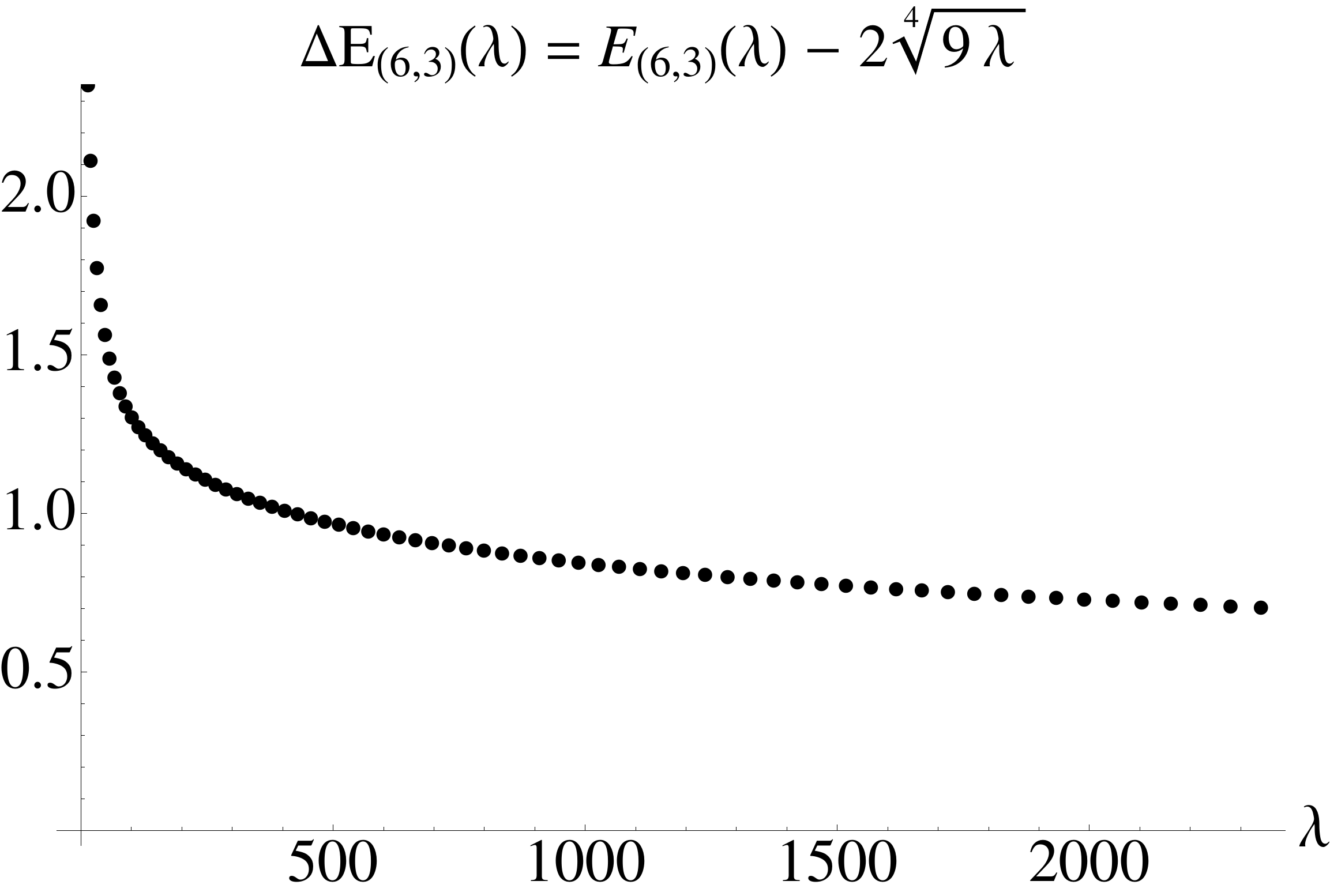}\qquad \includegraphics*[width=0.465\textwidth]{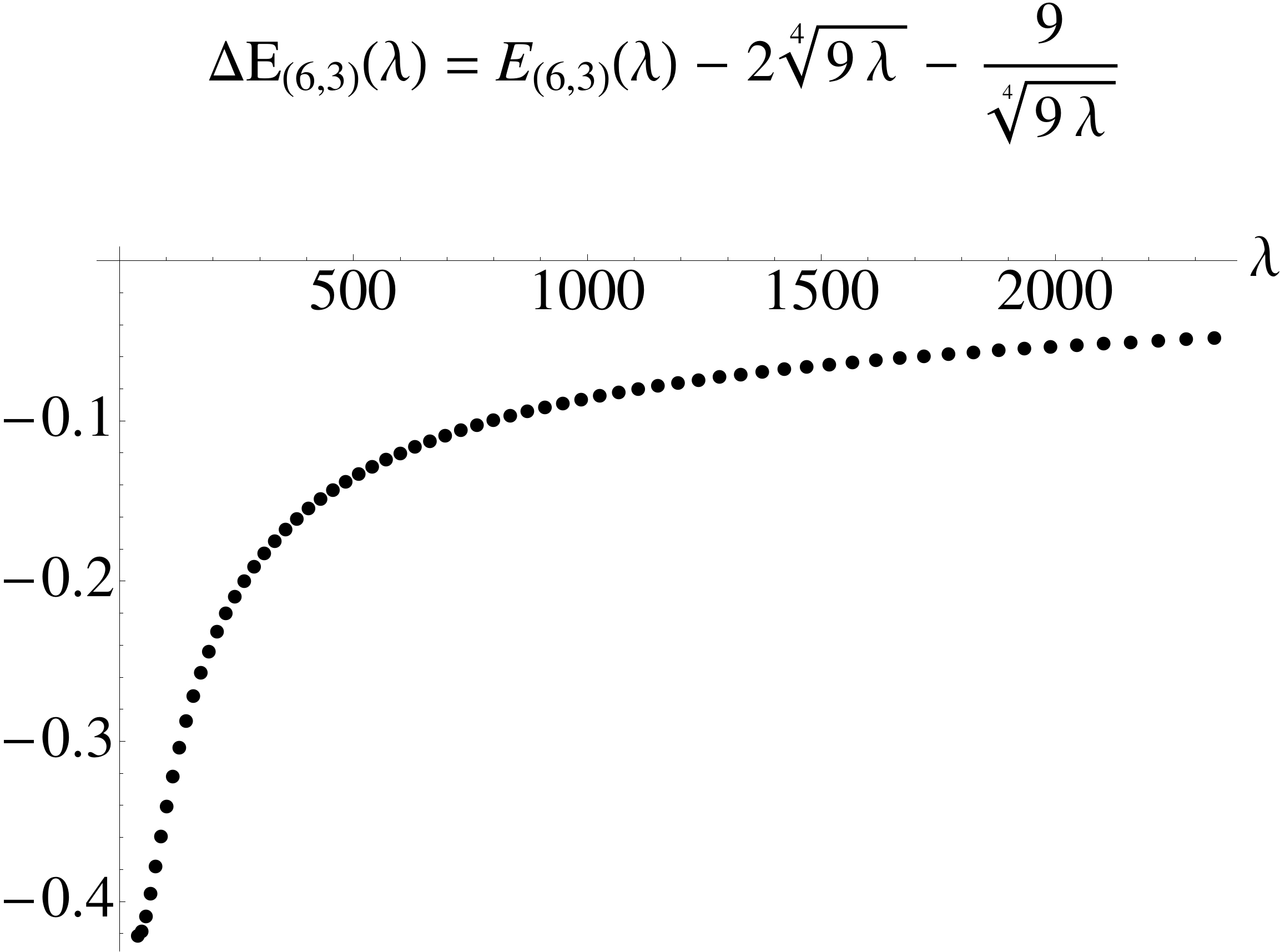}
\end{center}
\caption{\smaller 
Black dots represent the difference between the numerical solution and conjectured 
asymptotic expansions.
}
\la{figJ6n3}
\end{figure}
Then setting $c_1 = 9$, one gets
{\smaller 
\bea\la{FitEJ6n3d}
\begin{array}{|c|c|c|}
\hline
g_0&\lam_0& {\rm Fit} \\\hline
 2.2 & 191.076 & 2 \sqrt[4]{\Lambda }+\frac{9}{\sqrt[4]{\Lambda }}-\frac{0.683102}{\sqrt{\Lambda }}-\frac{91.8011}{\Lambda
   ^{3/4}}+\frac{133.416}{\Lambda }+\frac{416.847}{\Lambda ^{5/4}} \\
 2.3 & 208.841 & 2 \sqrt[4]{\Lambda }+\frac{9}{\sqrt[4]{\Lambda }}-\frac{0.646888}{\sqrt{\Lambda }}-\frac{92.7809}{\Lambda
   ^{3/4}}+\frac{142.105}{\Lambda }+\frac{391.588}{\Lambda ^{5/4}} \\
 2.4 & 227.396 & 2 \sqrt[4]{\Lambda }+\frac{9}{\sqrt[4]{\Lambda }}-\frac{0.641557}{\sqrt{\Lambda }}-\frac{92.9268}{\Lambda
   ^{3/4}}+\frac{143.414}{\Lambda }+\frac{387.73}{\Lambda ^{5/4}} \\
 2.5 & 246.74 & 2 \sqrt[4]{\Lambda }+\frac{9}{\sqrt[4]{\Lambda }}-\frac{0.659812}{\sqrt{\Lambda }}-\frac{92.422}{\Lambda
   ^{3/4}}+\frac{138.829}{\Lambda }+\frac{401.412}{\Lambda ^{5/4}}    \\\hline
   \end{array}~~~~
\eea
}Even though the coefficient $c_2$ is not really small  the contribution of the corresponding term is much smaller than the contribution of the next term, and we believe that this supports $c_2=0$. 
Setting $c_{2k}=0$ one gets the following fitting
{\smaller 
\bea\la{FitEJ6n3e}
\begin{array}{|c|c|c|}
\hline
g_0&\lam_0& {\rm Fit} \\\hline
2.2 & 191.076 & 2.00009 \sqrt[4]{\Lambda }+\frac{8.93337}{\sqrt[4]{\Lambda }}-\frac{85.2279}{\Lambda
   ^{3/4}}+\frac{1023.82}{\Lambda ^{5/4}}-\frac{3934.94}{\Lambda ^{7/4}} \\
 2.3 & 208.841 & 2.00006 \sqrt[4]{\Lambda }+\frac{8.94417}{\sqrt[4]{\Lambda }}-\frac{86.5518}{\Lambda
   ^{3/4}}+\frac{1092.42}{\Lambda ^{5/4}}-\frac{5204.87}{\Lambda ^{7/4}} \\
 2.4 & 227.396 & 2.00004 \sqrt[4]{\Lambda }+\frac{8.95121}{\sqrt[4]{\Lambda }}-\frac{87.4341}{\Lambda
   ^{3/4}}+\frac{1139.29}{\Lambda ^{5/4}}-\frac{6097.}{\Lambda ^{7/4}} \\
 2.5 & 246.74 & 2.00003 \sqrt[4]{\Lambda }+\frac{8.95445}{\sqrt[4]{\Lambda }}-\frac{87.8483}{\Lambda
   ^{3/4}}+\frac{1161.82}{\Lambda ^{5/4}}-\frac{6537.3}{\Lambda ^{7/4}}
    \\\hline
   \end{array}~~~~
\eea
}which agrees very well with $c_1=9$
but it shows that the coefficient $c_3$ is different from  the conjectured value
$c_3=-\frac{87}{8}-81 \zeta (3)\approx -108.242$.
Setting $c_{-1}=2, c_1=9$
{\smaller 
\bea\la{FitEJ6n3f}
\begin{array}{|c|c|c|}
\hline
g_0&\lam_0& {\rm Fit} \\\hline
2.2 & 191.076 & 2 \sqrt[4]{\Lambda }+\frac{9}{\sqrt[4]{\Lambda }}-\frac{96.7473}{\Lambda ^{3/4}}+\frac{1737.24}{\Lambda
   ^{5/4}}-\frac{18361.}{\Lambda ^{7/4}} \\
 2.3 & 208.841 & 2 \sqrt[4]{\Lambda }+\frac{9}{\sqrt[4]{\Lambda }}-\frac{96.9946}{\Lambda ^{3/4}}+\frac{1774.91}{\Lambda
   ^{5/4}}-\frac{19670.4}{\Lambda ^{7/4}} \\
 2.4 & 227.396 & 2 \sqrt[4]{\Lambda }+\frac{9}{\sqrt[4]{\Lambda }}-\frac{97.2184}{\Lambda ^{3/4}}+\frac{1809.99}{\Lambda
   ^{5/4}}-\frac{20931.5}{\Lambda ^{7/4}} \\
 2.5 & 246.74 & 2 \sqrt[4]{\Lambda }+\frac{9}{\sqrt[4]{\Lambda }}-\frac{97.4245}{\Lambda ^{3/4}}+\frac{1843.16}{\Lambda
   ^{5/4}}-\frac{22163.}{\Lambda ^{7/4}}
   \\\hline
   \end{array}~~~~
\eea
}does not help and one has to conclude that the formula conjectured in  \cite{Gr11} is correct only for the $n=1$ and $n=2$ operators. It is not really surprising because \cite{Gr11} used $c_1=10$ while the mirror TBA predicts $c_1=9$.  

In Figure \ref{figJ6n3} we plot the difference between the numerical solution and its large $\lam$ asymptotics 
$2(9\lam)^{1/4}$ and $2(9\lam)^{1/4}+{9\ov(9\lam)^{1/4}}$.

\subsection{$J=7,n=3$ operator}

In the table \eqref{E73data} we present the results of the computation of the energy of the $J=7,n=3$ state.
Fitting the data in the interval $g\in [g_0, 7.7]$ we get 
{\smaller 
\bea\la{FitEJ7n3a}
\begin{array}{|c|c|c|}
\hline
g_0&\lam_0& {\rm Fit} \\\hline
 2.2 & 191.076 & 2.02727 \sqrt[4]{\Lambda }-1.46037+\frac{44.517}{\sqrt[4]{\Lambda }}-\frac{377.119}{\sqrt{\Lambda
   }}+\frac{2352.97}{\Lambda ^{3/4}}-\frac{8336.68}{\Lambda }+\frac{12906.3}{\Lambda ^{5/4}} \\
 2.3 & 208.841 & 2.03163 \sqrt[4]{\Lambda }-1.70019+\frac{49.9676}{\sqrt[4]{\Lambda }}-\frac{442.652}{\sqrt{\Lambda
   }}+\frac{2792.56}{\Lambda ^{3/4}}-\frac{9896.42}{\Lambda }+\frac{15193.5}{\Lambda ^{5/4}} \\
 2.4 & 227.396 & 2.03791 \sqrt[4]{\Lambda }-2.04842+\frac{57.9602}{\sqrt[4]{\Lambda }}-\frac{539.755}{\sqrt{\Lambda
   }}+\frac{3451.11}{\Lambda ^{3/4}}-\frac{12260.3}{\Lambda }+\frac{18702.2}{\Lambda ^{5/4}} \\
 2.5 & 246.74 & 2.04761 \sqrt[4]{\Lambda }-2.59153+\frac{70.5447}{\sqrt[4]{\Lambda }}-\frac{694.179}{\sqrt{\Lambda
   }}+\frac{4509.47}{\Lambda ^{3/4}}-\frac{16101.5}{\Lambda }+\frac{24470.}{\Lambda ^{5/4}} \\\hline
   \end{array}~~~~
\eea
}where $\Lambda = n^2\lambda = 9\lambda$.
One sees that $c_{-1}$ is close to 2, and 
fixing  $c_{-1}= 2$ one gets
{\smaller 
\bea\la{FitEJ7n3b}
\begin{array}{|c|c|c|}
\hline
g_0&\lam_0& {\rm Fit} \\\hline
 2.2 & 191.076 & 2 \sqrt[4]{\Lambda }-0.00494946+\frac{12.4534}{\sqrt[4]{\Lambda }}-\frac{3.87439}{\sqrt{\Lambda
   }}-\frac{68.4585}{\Lambda ^{3/4}}-\frac{35.0256}{\Lambda }+\frac{1153.99}{\Lambda ^{5/4}} \\
 2.3 & 208.841 & 2 \sqrt[4]{\Lambda }+0.00564328+\frac{11.9695}{\sqrt[4]{\Lambda }}+\frac{4.88171}{\sqrt{\Lambda
   }}-\frac{146.9}{\Lambda ^{3/4}}+\frac{312.873}{\Lambda }+\frac{542.839}{\Lambda ^{5/4}} \\
 2.4 & 227.396 & 2 \sqrt[4]{\Lambda }+0.0162742+\frac{11.4791}{\sqrt[4]{\Lambda }}+\frac{13.8469}{\sqrt{\Lambda
   }}-\frac{228.106}{\Lambda ^{3/4}}+\frac{677.277}{\Lambda }-\frac{105.285}{\Lambda ^{5/4}} \\
 2.5 & 246.74 & 2 \sqrt[4]{\Lambda }+0.0264211+\frac{11.0067}{\sqrt[4]{\Lambda }}+\frac{22.5695}{\sqrt{\Lambda
   }}-\frac{307.95}{\Lambda ^{3/4}}+\frac{1039.59}{\Lambda }-\frac{757.345}{\Lambda ^{5/4}}
  \\\hline
   \end{array}~~~~
\eea
}Since the coefficient $c_0$ becomes small we set it to 0: $c_0=0$
{\smaller 
\bea\la{FitEJ7n3c}
\begin{array}{|c|c|c|}
\hline
g_0&\lam_0& {\rm Fit} \\\hline
2.2 & 191.076 & 2 \sqrt[4]{\Lambda }+\frac{12.2351}{\sqrt[4]{\Lambda }}-\frac{0.065948}{\sqrt{\Lambda
   }}-\frac{101.3}{\Lambda ^{3/4}}+\frac{104.997}{\Lambda }+\frac{917.781}{\Lambda ^{5/4}} \\
 2.3 & 208.841 & 2 \sqrt[4]{\Lambda }+\frac{12.2212}{\sqrt[4]{\Lambda }}+\frac{0.439529}{\sqrt{\Lambda
   }}-\frac{108.11}{\Lambda ^{3/4}}+\frac{145.265}{\Lambda }+\frac{829.583}{\Lambda ^{5/4}} \\
 2.4 & 227.396 & 2 \sqrt[4]{\Lambda }+\frac{12.2126}{\sqrt[4]{\Lambda }}+\frac{0.754727}{\sqrt{\Lambda
   }}-\frac{112.403}{\Lambda ^{3/4}}+\frac{170.959}{\Lambda }+\frac{772.582}{\Lambda ^{5/4}} \\
 2.5 & 246.74 & 2 \sqrt[4]{\Lambda }+\frac{12.2096}{\sqrt[4]{\Lambda }}+\frac{0.866542}{\sqrt{\Lambda
   }}-\frac{113.942}{\Lambda ^{3/4}}+\frac{180.276}{\Lambda }+\frac{751.66}{\Lambda ^{5/4}}
    \\\hline
   \end{array}~~~~
\eea
}\begin{figure}[t]
\begin{center}
\includegraphics*[width=0.475\textwidth]{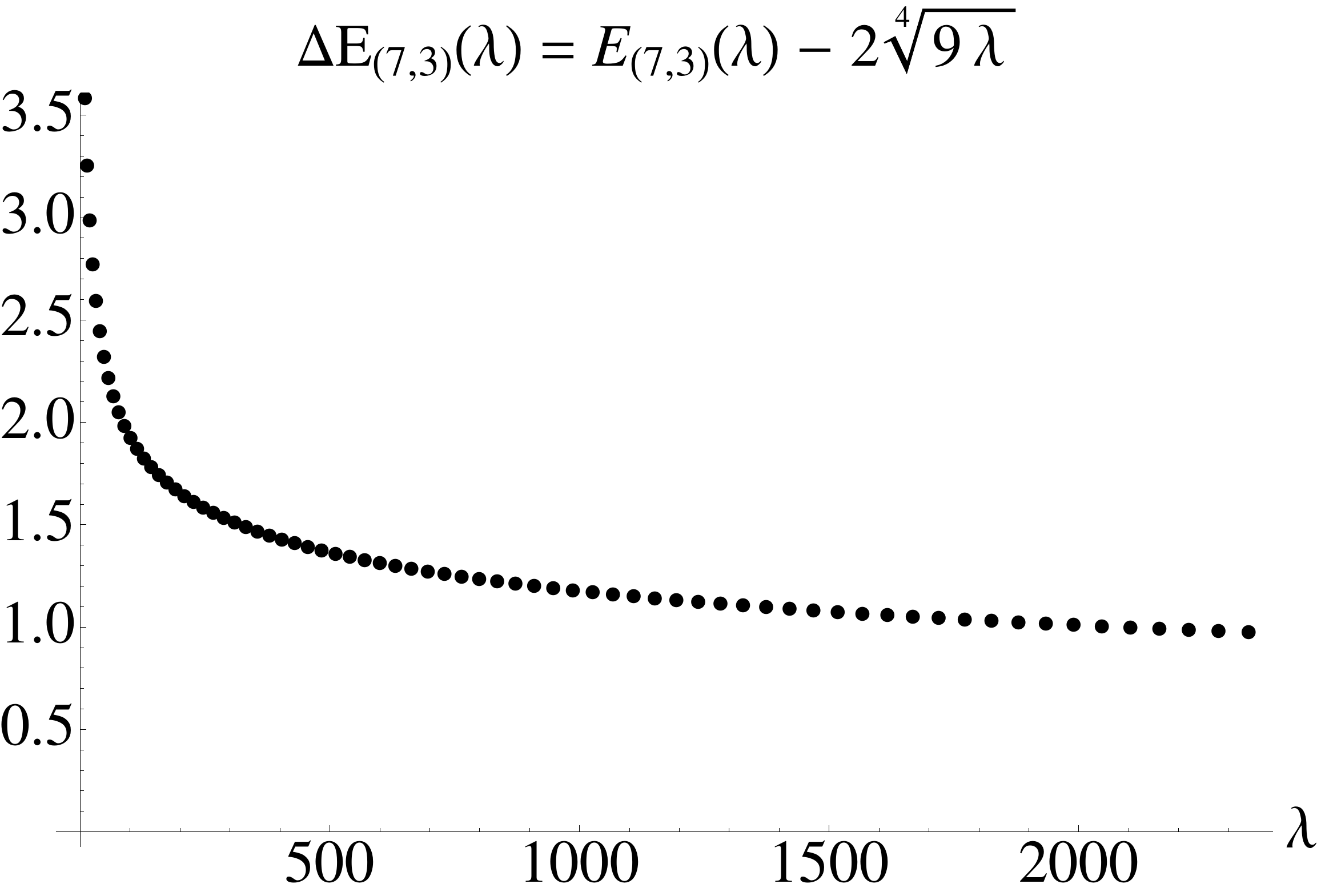}\qquad \includegraphics*[width=0.465\textwidth]{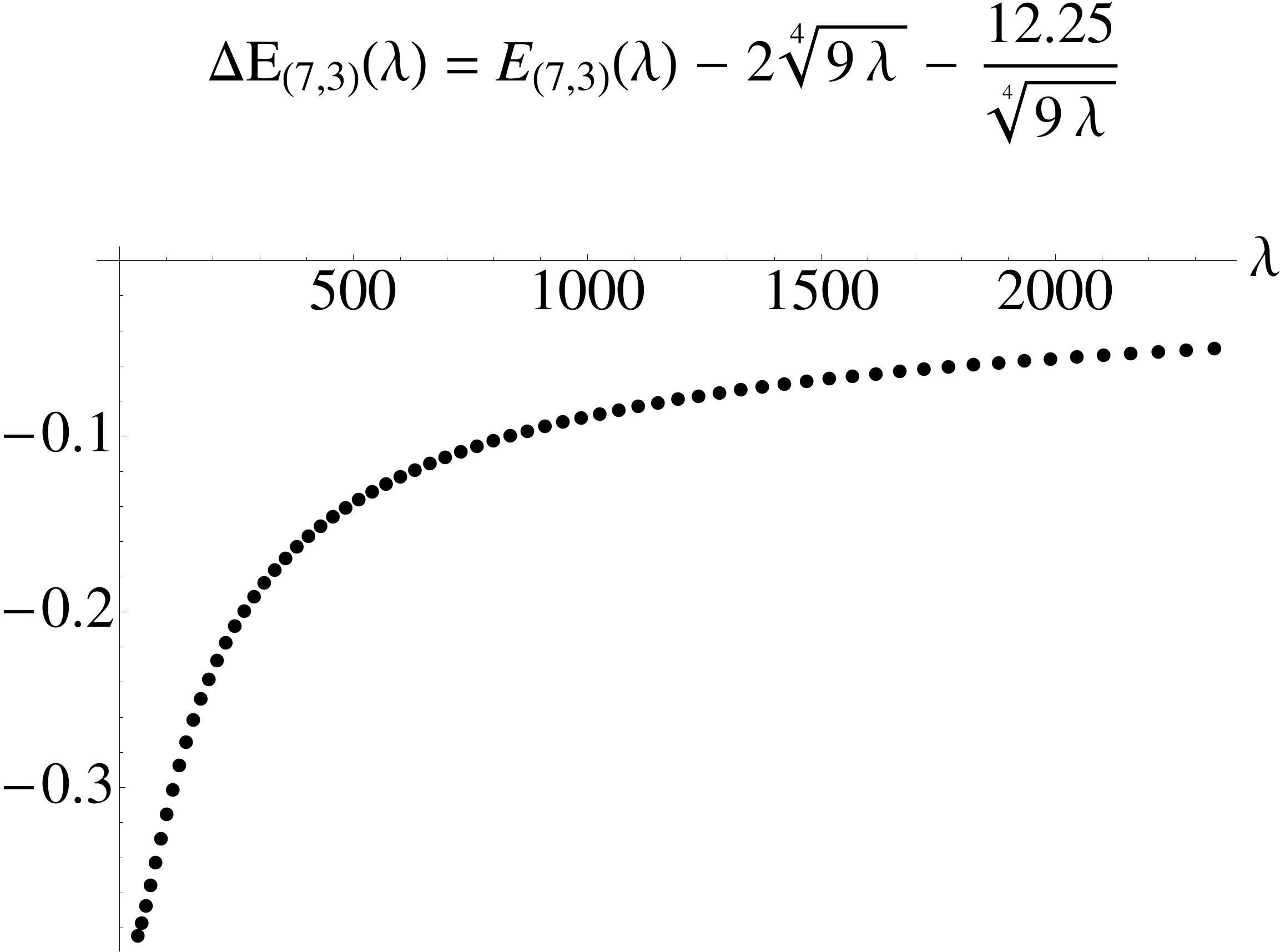}
\end{center}
\caption{\smaller 
Black dots represent the difference between the numerical solution and conjectured 
asymptotic expansions.
}
\la{figJ7n3}
\end{figure}This fitting shows that $c_1\approx 12.25$ in agreement with the  formula $c_1(J,2)=J^2/4$ we obtained analyzing  the $J=6,n=3$ state. 
Then setting $c_1 = 12.25$, one gets
{\smaller 
\bea\la{FitEJ7n3d}
\begin{array}{|c|c|c|}
\hline
g_0&\lam_0& {\rm Fit} \\\hline
 2.2 & 191.076 & 2 \sqrt[4]{\Lambda }+\frac{12.25}{\sqrt[4]{\Lambda }}-\frac{0.586725}{\sqrt{\Lambda
   }}-\frac{94.5925}{\Lambda ^{3/4}}+\frac{67.1513}{\Lambda }+\frac{996.766}{\Lambda ^{5/4}} \\
 2.3 & 208.841 & 2 \sqrt[4]{\Lambda }+\frac{12.25}{\sqrt[4]{\Lambda }}-\frac{0.577563}{\sqrt{\Lambda
   }}-\frac{94.8404}{\Lambda ^{3/4}}+\frac{69.3494}{\Lambda }+\frac{990.376}{\Lambda ^{5/4}} \\
 2.4 & 227.396 & 2 \sqrt[4]{\Lambda }+\frac{12.25}{\sqrt[4]{\Lambda }}-\frac{0.57989}{\sqrt{\Lambda
   }}-\frac{94.7767}{\Lambda ^{3/4}}+\frac{68.7779}{\Lambda }+\frac{992.06}{\Lambda ^{5/4}} \\
 2.5 & 246.74 & 2 \sqrt[4]{\Lambda }+\frac{12.25}{\sqrt[4]{\Lambda }}-\frac{0.59125}{\sqrt{\Lambda
   }}-\frac{94.4626}{\Lambda ^{3/4}}+\frac{65.9249}{\Lambda }+\frac{1000.57}{\Lambda ^{5/4}}
    \\\hline
   \end{array}~~~~
\eea
}Since the contribution of the $c_2$  term is much smaller than the contribution of the next term the fitting supports $c_2=0$. 
Setting $c_{2k}=0$ one gets the following fitting
{\smaller 
\bea\la{FitEJ7n3e}
\begin{array}{|c|c|c|}
\hline
g_0&\lam_0& {\rm Fit} \\\hline
2.2 & 191.076 & 2.00005 \sqrt[4]{\Lambda }+\frac{12.2061}{\sqrt[4]{\Lambda }}-\frac{93.0595}{\Lambda
   ^{3/4}}+\frac{1340.05}{\Lambda ^{5/4}}-\frac{2462.57}{\Lambda ^{7/4}} \\
 2.3 & 208.841 & 2.00004 \sqrt[4]{\Lambda }+\frac{12.2106}{\sqrt[4]{\Lambda }}-\frac{93.6089}{\Lambda
   ^{3/4}}+\frac{1368.52}{\Lambda ^{5/4}}-\frac{2989.6}{\Lambda ^{7/4}} \\
 2.4 & 227.396 & 2.00003 \sqrt[4]{\Lambda }+\frac{12.2136}{\sqrt[4]{\Lambda }}-\frac{93.9924}{\Lambda
   ^{3/4}}+\frac{1388.89}{\Lambda ^{5/4}}-\frac{3377.36}{\Lambda ^{7/4}} \\
 2.5 & 246.74 & 2.00003 \sqrt[4]{\Lambda }+\frac{12.2152}{\sqrt[4]{\Lambda }}-\frac{94.1927}{\Lambda
   ^{3/4}}+\frac{1399.79}{\Lambda ^{5/4}}-\frac{3590.34}{\Lambda ^{7/4}}
    \\\hline
   \end{array}~~~~
\eea
}which agrees with $c_1=12.25$
but disagrees with the conjectured value of $c_3$: 

\noindent $c_3=-\frac{1489}{64}-81 \zeta (3)\approx -120.632$.
Finally setting $c_{-1}=2, c_1=12.25$ one gets
{\smaller 
\bea\la{FitEJ7n3f}
\begin{array}{|c|c|c|}
\hline
g_0&\lam_0& {\rm Fit} \\\hline
2.2 & 191.076 & 2 \sqrt[4]{\Lambda }+\frac{12.25}{\sqrt[4]{\Lambda }}-\frac{100.961}{\Lambda
   ^{3/4}}+\frac{1838.01}{\Lambda ^{5/4}}-\frac{12626.6}{\Lambda ^{7/4}} \\
 2.3 & 208.841 & 2 \sqrt[4]{\Lambda }+\frac{12.25}{\sqrt[4]{\Lambda }}-\frac{101.136}{\Lambda
   ^{3/4}}+\frac{1864.62}{\Lambda ^{5/4}}-\frac{13551.4}{\Lambda ^{7/4}} \\
 2.4 & 227.396 & 2 \sqrt[4]{\Lambda }+\frac{12.25}{\sqrt[4]{\Lambda }}-\frac{101.298}{\Lambda
   ^{3/4}}+\frac{1890.13}{\Lambda ^{5/4}}-\frac{14468.5}{\Lambda ^{7/4}} \\
 2.5 & 246.74 & 2 \sqrt[4]{\Lambda }+\frac{12.25}{\sqrt[4]{\Lambda }}-\frac{101.452}{\Lambda
   ^{3/4}}+\frac{1914.79}{\Lambda ^{5/4}}-\frac{15384.2}{\Lambda ^{7/4}}
    \\\hline
   \end{array}~~~~
\eea
}In Figure \ref{figJ7n3} we plot the difference between the numerical solution and its large $\lam$ asymptotics 
$2(9\lam)^{1/4}$ and $2(9\lam)^{1/4}+{12.25\ov(9\lam)^{1/4}}$.

\subsection{$J=8,n=4$ operator}

In the table \eqref{E84data} we present the results of the computation of the energy of the $J=8,n=4$ state. The precision for this operator is higher than for the other operators and should be about $0.00002$. 
Fitting the data in the interval $g\in [g_0, 7.7]$ we get 
{\smaller 
\bea\la{FitEJ8n4a}
\begin{array}{|c|c|c|}
\hline
g_0&\lam_0& {\rm Fit} \\\hline
 2.4 & 227.396 & 2.02587 \sqrt[4]{\Lambda }-1.71713+\frac{61.1202}{\sqrt[4]{\Lambda }}-\frac{685.038}{\sqrt{\Lambda
   }}+\frac{5322.77}{\Lambda ^{3/4}}-\frac{23175.3}{\Lambda }+\frac{43004.2}{\Lambda ^{5/4}} \\
 2.5 & 246.74 & 2.0207 \sqrt[4]{\Lambda }-1.38273+\frac{52.1731}{\sqrt[4]{\Lambda }}-\frac{558.265}{\sqrt{\Lambda
   }}+\frac{4319.5}{\Lambda ^{3/4}}-\frac{18970.8}{\Lambda }+\frac{35714.1}{\Lambda ^{5/4}} \\
 2.6 & 266.874 & 2.02039 \sqrt[4]{\Lambda }-1.36242+\frac{51.6249}{\sqrt[4]{\Lambda }}-\frac{550.422}{\sqrt{\Lambda
   }}+\frac{4256.81}{\Lambda ^{3/4}}-\frac{18705.2}{\Lambda }+\frac{35248.6}{\Lambda ^{5/4}} \\
 2.7 & 287.798 & 2.0272 \sqrt[4]{\Lambda }-1.81055+\frac{63.8305}{\sqrt[4]{\Lambda }}-\frac{726.645}{\sqrt{\Lambda
   }}+\frac{5679.17}{\Lambda ^{3/4}}-\frac{24790.5}{\Lambda }+\frac{46029.7}{\Lambda ^{5/4}}\\\hline
   \end{array}~~~~
\eea
}where $\Lambda = n^2\lambda = 16\lambda$.
One sees that $c_{-1}$ is close to 2, and 
fixing  $c_{-1}= 2$ one gets
{\smaller 
\bea\la{FitEJ8n4b}
\begin{array}{|c|c|c|}
\hline
g_0&\lam_0& {\rm Fit} \\\hline
 2.4 & 227.396 & 2 \sqrt[4]{\Lambda }-0.0901497+\frac{18.8267}{\sqrt[4]{\Lambda }}-\frac{103.386}{\sqrt{\Lambda
   }}+\frac{859.119}{\Lambda ^{3/4}}-\frac{5051.13}{\Lambda }+\frac{12580.9}{\Lambda ^{5/4}} \\
 2.5 & 246.74 & 2 \sqrt[4]{\Lambda }-0.0684271+\frac{17.659}{\sqrt[4]{\Lambda }}-\frac{78.4877}{\sqrt{\Lambda
   }}+\frac{595.952}{\Lambda ^{3/4}}-\frac{3672.19}{\Lambda }+\frac{9715.32}{\Lambda ^{5/4}} \\
 2.6 & 266.874 & 2 \sqrt[4]{\Lambda }-0.0558615+\frac{16.9775}{\sqrt[4]{\Lambda }}-\frac{63.8179}{\sqrt{\Lambda
   }}+\frac{439.324}{\Lambda ^{3/4}}-\frac{2842.69}{\Lambda }+\frac{7972.04}{\Lambda ^{5/4}} \\
 2.7 & 287.798 & 2 \sqrt[4]{\Lambda }-0.0516738+\frac{16.7484}{\sqrt[4]{\Lambda }}-\frac{58.8418}{\sqrt{\Lambda
   }}+\frac{385.679}{\Lambda ^{3/4}}-\frac{2555.67}{\Lambda }+\frac{7362.33}{\Lambda ^{5/4}}
  \\\hline
   \end{array}~~~~
\eea
}Since the coefficient $c_0$ becomes small we set it to 0: $c_0=0$
{\smaller 
\bea\la{FitEJ8n4c}
\begin{array}{|c|c|c|}
\hline
g_0&\lam_0& {\rm Fit} \\\hline
 2.4 & 227.396 & 2 \sqrt[4]{\Lambda }+\frac{14.1353}{\sqrt[4]{\Lambda }}-\frac{6.68802}{\sqrt{\Lambda
   }}-\frac{127.649}{\Lambda ^{3/4}}-\frac{64.996}{\Lambda }+\frac{2598.41}{\Lambda ^{5/4}} \\
 2.5 & 246.74 & 2 \sqrt[4]{\Lambda }+\frac{14.0619}{\sqrt[4]{\Lambda }}-\frac{3.54403}{\sqrt{\Lambda
   }}-\frac{177.626}{\Lambda ^{3/4}}+\frac{284.294}{\Lambda }+\frac{1692.72}{\Lambda ^{5/4}} \\
 2.6 & 266.874 & 2 \sqrt[4]{\Lambda }+\frac{14.0123}{\sqrt[4]{\Lambda }}-\frac{1.39941}{\sqrt{\Lambda
   }}-\frac{212.063}{\Lambda ^{3/4}}+\frac{527.596}{\Lambda }+\frac{1054.52}{\Lambda ^{5/4}} \\
 2.7 & 287.798 & 2 \sqrt[4]{\Lambda }+\frac{13.9798}{\sqrt[4]{\Lambda }}+\frac{0.0198075}{\sqrt{\Lambda
   }}-\frac{235.074}{\Lambda ^{3/4}}+\frac{691.858}{\Lambda }+\frac{618.868}{\Lambda ^{5/4}}
    \\\hline
   \end{array}~~~~
\eea
}\begin{figure}[t]
\begin{center}
\includegraphics*[width=0.475\textwidth]{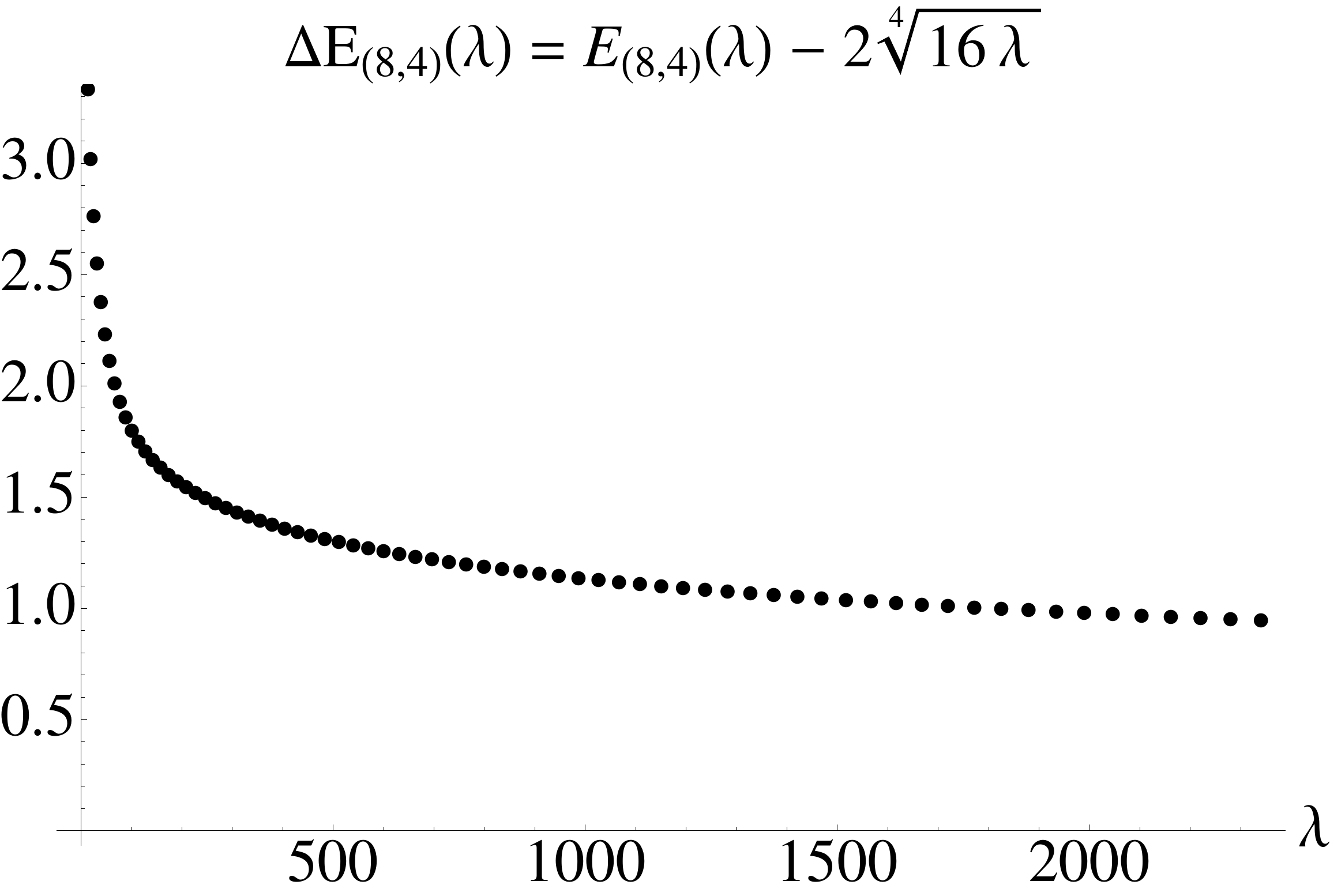}\qquad \includegraphics*[width=0.45\textwidth]{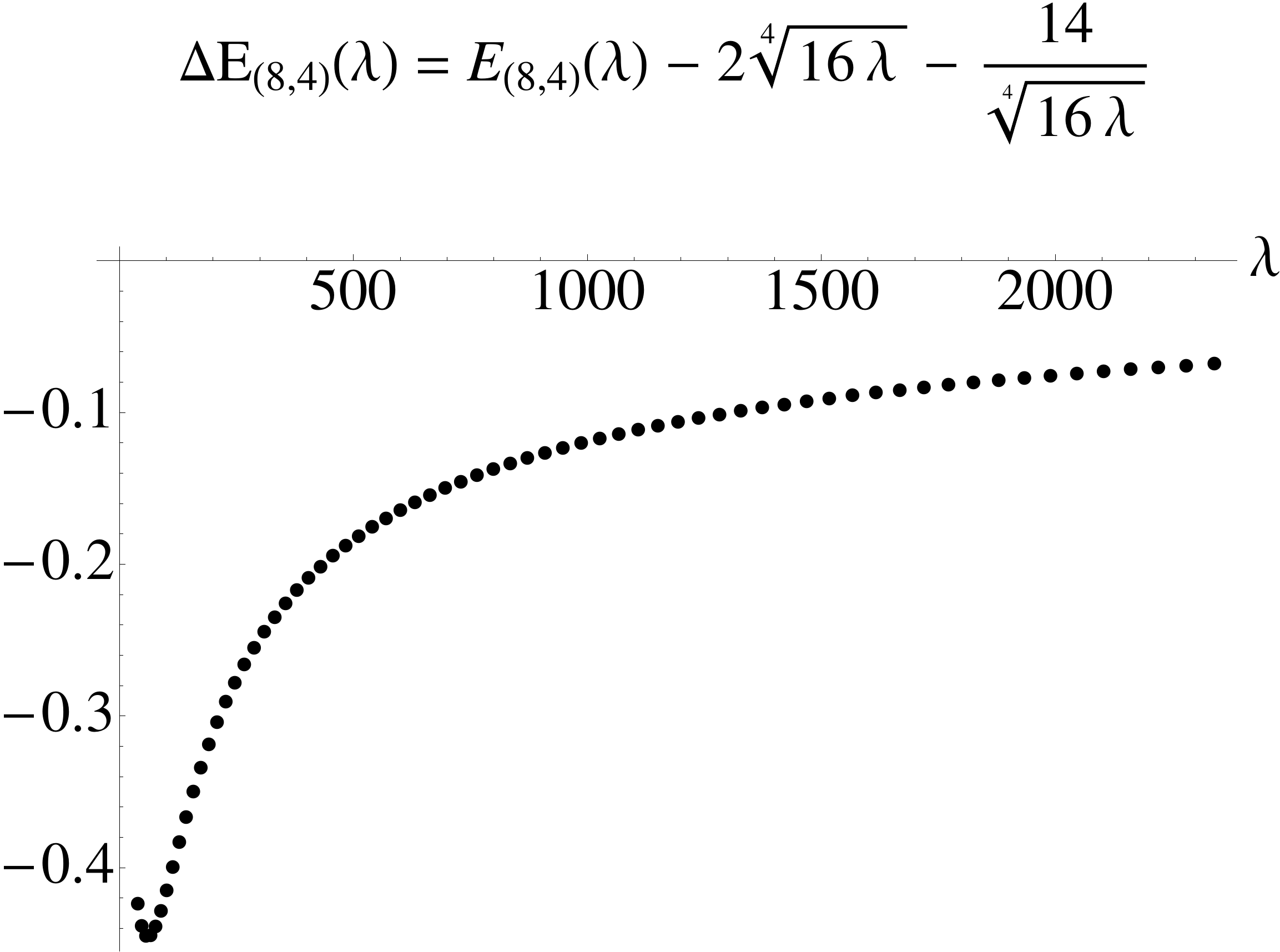}
\end{center}
\caption{\smaller 
Black dots represent the difference between the numerical solution and conjectured 
asymptotic expansions.
}
\la{figJ8n4}
\end{figure}This fitting shows that $c_1\approx 14$ which suggests the  formula $c_1(J,2)=J^2/4-2$. 
Then setting $c_1 = 14$, one gets
{\smaller 
\bea\la{FitEJ8n4d}
\begin{array}{|c|c|c|}
\hline
g_0&\lam_0& {\rm Fit} \\\hline
 2.4 & 227.396 & 2 \sqrt[4]{\Lambda }+\frac{14}{\sqrt[4]{\Lambda }}-\frac{1.11641}{\sqrt{\Lambda
   }}-\frac{212.617}{\Lambda ^{3/4}}+\frac{503.771}{\Lambda }+\frac{1187.75}{\Lambda ^{5/4}} \\
 2.5 & 246.74 & 2 \sqrt[4]{\Lambda }+\frac{14}{\sqrt[4]{\Lambda }}-\frac{0.966912}{\sqrt{\Lambda
   }}-\frac{217.391}{\Lambda ^{3/4}}+\frac{553.832}{\Lambda }+\frac{1015.24}{\Lambda ^{5/4}} \\
 2.6 & 266.874 & 2 \sqrt[4]{\Lambda }+\frac{14}{\sqrt[4]{\Lambda }}-\frac{0.881553}{\sqrt{\Lambda
   }}-\frac{220.144}{\Lambda ^{3/4}}+\frac{583.027}{\Lambda }+\frac{913.412}{\Lambda ^{5/4}} \\
 2.7 & 287.798 & 2 \sqrt[4]{\Lambda }+\frac{14}{\sqrt[4]{\Lambda }}-\frac{0.839038}{\sqrt{\Lambda
   }}-\frac{221.528}{\Lambda ^{3/4}}+\frac{597.866}{\Lambda }+\frac{861.055}{\Lambda ^{5/4}}
    \\\hline
   \end{array}~~~~
\eea
}Since the contribution of the $c_2$  term is much smaller than the contribution of the next term the fitting supports $c_2=0$. 
Setting $c_{2k}=0$ one gets the following fitting
{\smaller 
\bea\la{FitEJ8n4e}
\begin{array}{|c|c|c|}
\hline
g_0&\lam_0& {\rm Fit} \\\hline
 2.4 & 227.396 & 2.00022 \sqrt[4]{\Lambda }+\frac{13.8411}{\sqrt[4]{\Lambda }}-\frac{180.225}{\Lambda
   ^{3/4}}+\frac{3375.94}{\Lambda ^{5/4}}-\frac{13371.7}{\Lambda ^{7/4}} \\
 2.5 & 246.74 & 2.00017 \sqrt[4]{\Lambda }+\frac{13.8627}{\sqrt[4]{\Lambda }}-\frac{183.91}{\Lambda
   ^{3/4}}+\frac{3643.1}{\Lambda ^{5/4}}-\frac{20334.4}{\Lambda ^{7/4}} \\
 2.6 & 266.874 & 2.00014 \sqrt[4]{\Lambda }+\frac{13.8782}{\sqrt[4]{\Lambda }}-\frac{186.603}{\Lambda
   ^{3/4}}+\frac{3842.8}{\Lambda ^{5/4}}-\frac{25671.9}{\Lambda ^{7/4}} \\
 2.7 & 287.798 & 2.00012 \sqrt[4]{\Lambda }+\frac{13.8893}{\sqrt[4]{\Lambda }}-\frac{188.555}{\Lambda
   ^{3/4}}+\frac{3990.66}{\Lambda ^{5/4}}-\frac{29719.7}{\Lambda ^{7/4}}
    \\\hline
   \end{array}~~~~
\eea
}which agrees with $c_1=14$.
The agreement becomes even better if one sets $c_{-1}=2$ 
{\smaller 
\bea\la{FitEJ8n4f}
\begin{array}{|c|c|c|}
\hline
g_0&\lam_0& {\rm Fit} \\\hline
 2.4 & 227.396 & 2 \sqrt[4]{\Lambda }+\frac{13.9364}{\sqrt[4]{\Lambda }}-\frac{195.091}{\Lambda
   ^{3/4}}+\frac{4356.69}{\Lambda ^{5/4}}-\frac{36551.8}{\Lambda ^{7/4}} \\
 2.5 & 246.74 & 2 \sqrt[4]{\Lambda }+\frac{13.9395}{\sqrt[4]{\Lambda }}-\frac{196.166}{\Lambda
   ^{3/4}}+\frac{4473.31}{\Lambda ^{5/4}}-\frac{40538.4}{\Lambda ^{7/4}} \\
 2.6 & 266.874 & 2 \sqrt[4]{\Lambda }+\frac{13.9418}{\sqrt[4]{\Lambda }}-\frac{196.979}{\Lambda
   ^{3/4}}+\frac{4563.51}{\Lambda ^{5/4}}-\frac{43705.8}{\Lambda ^{7/4}} \\
 2.7 & 287.798 & 2 \sqrt[4]{\Lambda }+\frac{13.9435}{\sqrt[4]{\Lambda }}-\frac{197.604}{\Lambda
   ^{3/4}}+\frac{4634.52}{\Lambda ^{5/4}}-\frac{46263.4}{\Lambda ^{7/4}}
    \\\hline
   \end{array}~~~~
\eea
}In Figure \ref{figJ8n4} we plot the difference between the numerical solution and its large $\lam$ asymptotics 
$2(16\lam)^{1/4}$ and $2(16\lam)^{1/4}+{14\ov(16\lam)^{1/4}}$.

\section*{Acknowledgements}
The author thanks Gleb Arutyunov for interesting
discussions and Dmitri Grigoriev for computer help.  This work 
was supported in part by the Science Foundation Ireland under
Grant No.  09/RFP/PHY2142.

\section{Numerical data}

Here we collect our numerical data for the energy of the two-paticle states or, equivalently, the conformal dimension of the dual ${\cal N}=4$ SYM operators
as a function of the effective string tension $g$ related to 't Hooft's coupling $\lam$ as $\lam=4\pi^2g^2$.  

An interested reader can download a Mathematica file with this data and the data for Bethe roots from the Arxiv page of this paper:  http://arXiv.org/abs/arXiv:12??.????

\subsection*{ $J=3,n=1$ operator}
{\smaller \bea\la{E31data}
\begin{array}{|c|c||}
\hline
g& {\rm E}_{(3,1)}\\\hline
0 & 5 \\
0.1 & 5.01985 \\
 0.2 & 5.07773 \\
 0.3 & 5.16917 \\
 0.4 & 5.28828 \\
 0.5 & 5.42891 \\
 0.6 & 5.5854 
 \\\hline
\end{array}
\begin{array}{c|c||}
\hline
g& {\rm E}_{(3,1)}\\\hline
 0.7 & 5.75288 \\
 0.8 & 5.92734 \\
 0.9 & 6.10559 \\
 1. & 6.2862 \\
 1.1 & 6.46621 \\
 1.2 & 6.64492 \\
 1.3 & 6.82157 
 \\\hline
\end{array}
\begin{array}{c|c||}
\hline
g& {\rm E}_{(3,1)}\\\hline
 1.4 & 6.9957 \\
 1.5 & 7.16701 \\
 1.6 & 7.33539 \\
 1.7 & 7.50075 \\
 1.8 & 7.66321 \\
 1.9 & 7.82271 \\
 2. & 7.97949 
 \\\hline
\end{array}
\begin{array}{c|c||}
\hline
g& {\rm E}_{(3,1)}\\\hline
 2.1 & 8.13347 \\
 2.2 & 8.28485 \\
 2.3 & 8.4337 \\
 2.4 & 8.58018 \\
 2.5 & 8.72423 \\
 2.6 & 8.86614 \\
 2.7 & 9.00592 
 \\\hline
\end{array}
\begin{array}{c|c||}
\hline
g& {\rm E}_{(3,1)}\\\hline
 2.8 & 9.14362 \\
 2.9 & 9.27936 \\
 3. & 9.41322 \\
 3.1 & 9.54522 \\
 3.2 & 9.67549 \\
 3.3 & 9.80404 \\
 3.4 & 9.93099 
 \\\hline
\end{array}
\begin{array}{c|c|}
\hline
g& {\rm E}_{(3,1)}\\\hline
 3.5 & 10.0564 \\
 3.6 & 10.1802 \\
 3.7 & 10.3025\\
 &\\
 &\\
 &\\
 &
 \\\hline
\end{array}
\eea
}

\subsection*{$J=4,n=1$ operator}
{\smaller \bea\la{E41data}
\begin{array}{|c|c||}
\hline
g& {\rm E}_{(4,1)}\\\hline
0 & 6 \\
0.1 & 6.01375 \\
 0.2 & 6.05416 \\
 0.3 & 6.11897 \\
 0.4 & 6.20502 \\
 0.5 & 6.30878 
 \\\hline
\end{array}
\begin{array}{c|c||}
\hline
g& {\rm E}_{(4,1)}\\\hline
 0.6 & 6.42681\\
 0.7 & 6.55594 \\
 0.8 & 6.6934 \\
 0.9 & 6.83685 \\
 1. & 6.98458 \\
 1.1 & 7.13475 
  \\\hline
\end{array}
\begin{array}{c|c||}
\hline
g& {\rm E}_{(4,1)}\\\hline
 1.2 & 7.28623 \\
 1.3 & 7.43822 \\
 1.4 & 7.58999 \\
 1.5 & 7.74085 \\
 1.6 & 7.89072 \\
 1.61 & 7.90572  
 \\\hline
\end{array}
\begin{array}{c|c||}
\hline
g& {\rm E}_{(4,1)}\\\hline
1.62 & 7.92062 \\
 1.63 & 7.93556 \\
 2.2 & 8.75551\\
 2.3 & 8.89485 \\
 2.4 & 9.03165 \\
 2.5 & 9.16659 
 \\\hline
\end{array}
\begin{array}{c|c||}
\hline
g& {\rm E}_{(4,1)}\\\hline
 2.6 & 9.29982 \\
 2.7 & 9.43134 \\
 2.8 & 9.56136 \\
 2.9 & 9.68974 \\
 3. & 9.81658 \\
 3.1 & 9.94203 
 \\\hline
\end{array}
\begin{array}{c|c|}
\hline
g& {\rm E}_{(4,1)}\\\hline
 3.2 & 10.0659 \\
 3.3 & 10.1884 \\
 3.4 & 10.3096\\
 &\\
 &\\
 &
 \\\hline
\end{array}
\eea
}

\subsection*{$J=4,n=2$ operator}

{\smaller \bea\la{E42data}
\begin{array}{|c|c||}
\hline
g& {\rm E}_{(4,2)} \\\hline
0 & 6 \\
 0.1 & 6.03583 \\
 0.2 & 6.13947 \\
 0.3 & 6.30103 \\
 0.4 & 6.50828 \\
 0.5 & 6.74951 \\
 0.6 & 7.01484 \\
 0.7 & 7.29635 \\
 0.8 & 7.58773 \\
 0.9 & 7.88397 \\
 1. & 8.18112 \\
 1.1 & 8.47607 \\
 1.2 & 8.7666 
   \\\hline
\end{array}
\begin{array}{c|c||}
\hline
g& {\rm E}_{(4,2)} \\\hline
 1.3 & 9.05125 \\
 1.4 & 9.32919 \\
 1.5 & 9.60001 \\
 1.6 & 9.8638 \\
 1.7 & 10.1208 \\
 1.8 & 10.3713 \\
 1.9 & 10.6157 \\
 2. & 10.8543 \\
 2.1 & 11.0876 \\
 2.2 & 11.3158 \\
 2.3 & 11.5393 \\
 2.4 & 11.7583 \\
 2.5 & 11.9732 
   \\\hline
\end{array}
\begin{array}{c|c||}
\hline
g& {\rm E}_{(4,2)} \\\hline
 2.6 & 12.1841 \\
 2.7 & 12.3912 \\
 2.8 & 12.5948 \\
 2.9 & 12.795 \\
 3. & 12.9921 \\
 3.1 & 13.1861 \\
 3.2 & 13.3771 \\
 3.3 & 13.5654 \\
 3.4 & 13.7511 \\
 3.5 & 13.9342 \\
 3.6 & 14.1148 \\
 3.7 & 14.2931 \\
 3.8 & 14.4691 
    \\\hline
\end{array}
\begin{array}{c|c||}
\hline
g& {\rm E}_{(4,2)} \\\hline
 3.9 & 14.643 \\
 4. & 14.8147 \\
 4.1 & 14.9845 \\
 4.2 & 15.1523 \\
 4.3 & 15.3182 \\
 4.4 & 15.4824 \\
 4.5 & 15.6447 \\
 4.6 & 15.8054 \\
 4.7 & 15.9644 \\
 4.8 & 16.1218 \\
 4.9 & 16.2777 \\
 5. & 16.4321 \\
 5.1 & 16.585 
   \\\hline
\end{array}
\begin{array}{c|c||}
\hline
g& {\rm E}_{(4,2)} \\\hline
 5.2 & 16.7364 \\
 5.3 & 16.8865 \\
 5.4 & 17.0353 \\
 5.5 & 17.1828 \\
 5.6 & 17.3289 \\
 5.7 & 17.4739 \\
 5.8 & 17.6176 \\
 5.9 & 17.7601 \\
 6. & 17.9016 \\
 6.1 & 18.0418 \\
 6.2 & 18.181 \\
 6.3 & 18.3191 \\
 6.4 & 18.4562
   \\\hline
\end{array}
\begin{array}{c|c|}
\hline
g& {\rm E}_{(4,2)} \\\hline
 6.5 & 18.5923 \\
 6.6 & 18.7273 \\
 6.7 & 18.8614 \\
 6.8 & 18.9945 \\
 6.9 & 19.1267 \\
 7. & 19.258 \\
 7.1 & 19.3883 \\
 7.2 & 19.5177 \\
 7.3 & 19.6464 \\
 7.4 & 19.7742 \\
 7.5 & 19.9012 \\
 7.6 & 20.0273 \\
 7.7 & 20.1526
   \\\hline
\end{array}
\eea
}

\subsection*{$J=5,n=2$ operator}
{\smaller \bea\la{E52data}
\begin{array}{|c|c||}
\hline
g& {\rm E}_{(5,2)}\\
\hline
  0 & 7\\
  0.1 & 7.02974 \\
 0.2 & 7.11606 \\
 0.3 & 7.25139 \\
 0.4 & 7.42611 \\
 0.5 & 7.63068 \\
 0.6 & 7.85687 \\
 0.7 & 8.09799 \\
 0.8 & 8.3488 \\
 0.9 & 8.60527 \\
 1. & 8.86434 \\
 1.1 & 9.12364 \\
 1.2 & 9.38146 
   \\\hline
\end{array}
\begin{array}{c|c||}
\hline
g& {\rm E}_{(5,2)}\\
\hline
 1.3 & 9.63657 \\
 1.4 & 9.88812 \\
 1.5 & 10.1356 \\
 1.6 & 10.3787 \\
 1.7 & 10.6173 \\
 1.8 & 10.8513 \\
 1.9 & 11.081 \\
 2. & 11.3063 \\
 2.1 & 11.5274 \\
 2.2 & 11.7445 \\
 2.3 & 11.9577 \\
 2.4 & 12.1673 \\
 2.5 & 12.3734 
  \\\hline
\end{array}
\begin{array}{c|c||}
\hline
g& {\rm E}_{(5,2)}\\
\hline
 2.6 & 12.576 \\
 2.7 & 12.7755 \\
 2.8 & 12.9719 \\
 2.9 & 13.1653 \\
 3. & 13.3559 \\
 3.1 & 13.5438 \\
 3.2 & 13.7291 \\
 3.3 & 13.912 \\
 3.4 & 14.0924 \\
 3.5 & 14.2705 \\
 3.6 & 14.4464 \\
 3.7 & 14.6201 \\
 3.8 & 14.7918  
  \\\hline
\end{array}
\begin{array}{c|c||}
\hline
g& {\rm E}_{(5,2)}\\
\hline
 3.9 & 14.9615 \\
 4. & 15.1292 \\
 4.1 & 15.2951 \\
 4.2 & 15.4592 \\
 4.3 & 15.6215 \\
 4.4 & 15.7822 \\
 4.5 & 15.9412 \\
 4.6 & 16.0986 \\
 4.7 & 16.2545 \\
 4.8 & 16.4089 \\
 4.9 & 16.5619 \\
 5. & 16.7134 \\
 5.1 & 16.8636
  \\\hline
\end{array}
\begin{array}{c|c||}
\hline
g& {\rm E}_{(5,2)}\\
\hline
 5.2 & 17.0124 \\
 5.3 & 17.1599 \\
 5.4 & 17.3061 \\
 5.5 & 17.4511 \\
 5.6 & 17.5949 \\
 5.7 & 17.7375 \\
 5.8 & 17.879 \\
 5.9 & 18.0194 \\
 6. & 18.1586 \\
 6.1 & 18.2968 \\
 6.2 & 18.434 \\
 6.3 & 18.5701 \\
 6.4 & 18.7052
  \\\hline
\end{array}
\begin{array}{c|c|}
\hline
g& {\rm E}_{(5,2)}\\
\hline
 6.5 & 18.8393 \\
 6.6 & 18.9725 \\
 6.7 & 19.1048 \\
 6.8 & 19.2361 \\
 6.9 & 19.3666 \\
 7. & 19.4962 \\
 7.1 & 19.6249 \\
 7.2 & 19.7527 \\
 7.3 & 19.8798 \\
 7.4 & 20.006 \\
 7.5 & 20.1314 \\
 7.6 & 20.2561 \\
 7.7 & 20.3799 
  \\\hline
\end{array}
\eea
}

\subsection*{$J=6,n=3$ operator}

{\smaller \bea\la{E63data}
\begin{array}{|c|c||}
\hline
g& {\rm E}_{(6,3)}\\\hline
0&8\\
0.1 & 8.03765 \\
 0.2 & 8.14655 \\
 0.3 & 8.31636 \\
 0.4 & 8.53443 \\
 0.5 & 8.78889 \\
 0.6 & 9.06998 \\
 0.7 & 9.37014 \\
 0.8 & 9.68362 \\
 0.9 & 10.006 \\
 1. & 10.3337 \\
 1.1 & 10.6639 \\
 1.2 & 10.9941 
  \\\hline
\end{array}
\begin{array}{c|c||}
\hline
g& {\rm E}_{(6,3)}\\\hline
 1.3 & 11.3223 \\
 1.4 & 11.6469 \\
 1.5 & 11.9667 \\
 1.6 & 12.2807 \\
 1.7 & 12.5884 \\
 1.8 & 12.8897 \\
 1.9 & 13.1845 \\
 2. & 13.4729 \\
 2.1 & 13.7553 \\
 2.2 & 14.0319 \\
 2.3 & 14.3029 \\
 2.4 & 14.5687 \\
 2.5 & 14.8296
  \\\hline
\end{array}
\begin{array}{c|c||}
\hline
g& {\rm E}_{(6,3)}\\\hline
 2.6 & 15.0857 \\
 2.7 & 15.3374 \\
 2.8 & 15.5848 \\
 2.9 & 15.8282 \\
 3. & 16.0677 \\
 3.1 & 16.3036 \\
 3.2 & 16.536 \\
 3.3 & 16.765 \\
 3.4 & 16.9909 \\
 3.5 & 17.2136 \\
 3.6 & 17.4335 \\
 3.7 & 17.6505 \\
 3.8 & 17.8647 
  \\\hline
\end{array}
\begin{array}{c|c||}
\hline
g& {\rm E}_{(6,3)}\\\hline
 3.9 & 18.0764 \\
 4. & 18.2855 \\
 4.1 & 18.4922 \\
 4.2 & 18.6966 \\
 4.3 & 18.8987 \\
 4.4 & 19.0986 \\
 4.5 & 19.2964 \\
 4.6 & 19.4921 \\
 4.7 & 19.6858 \\
 4.8 & 19.8776 \\
 4.9 & 20.0675 \\
 5. & 20.2556 \\
 5.1 & 20.4419 
  \\\hline
\end{array}
\begin{array}{c|c||}
\hline
g& {\rm E}_{(6,3)}\\\hline
 5.2 & 20.6266 \\
 5.3 & 20.8094 \\
 5.4 & 20.9908 \\
 5.5 & 21.1705 \\
 5.6 & 21.3487 \\
 5.7 & 21.5254 \\
 5.8 & 21.7007 \\
 5.9 & 21.8745 \\
 6. & 22.0469 \\
 6.1 & 22.2179 \\
 6.2 & 22.3876 \\
 6.3 & 22.5561 \\
 6.4 & 22.7232
  \\\hline
\end{array}
\begin{array}{c|c|}
\hline
g& {\rm E}_{(6,3)}\\\hline
 6.5 & 22.8892 \\
 6.6 & 23.0539 \\
 6.7 & 23.2174 \\
 6.8 & 23.3798 \\
 6.9 & 23.541 \\
 7. & 23.7011 \\
 7.1 & 23.8602 \\
 7.2 & 24.0182 \\
 7.3 & 24.1751 \\
 7.4 & 24.331 \\
 7.5 & 24.4859 \\
 7.6 & 24.6398 \\
 7.7 & 24.7927
  \\\hline
\end{array}
\eea
}

\subsection*{$J=7,n=3$ operator}

{\smaller \bea\la{E73data}
\begin{array}{|c|c||}
\hline
g& {\rm E}_{(7,3)}\\\hline
0&9\\
 0.1 & 9.03383 \\
 0.2 & 9.13189 \\
 0.3 & 9.28531 \\
 0.4 & 9.483 \\
 0.5 & 9.71427 \\
 0.6 & 9.97009 \\
 0.7 & 10.2434 \\
 0.8 & 10.5286 \\
 0.9 & 10.8216 \\
 1. & 11.1193 \\
 1.1 & 11.4192 \\
 1.2 & 11.7193 
  \\\hline
\end{array}
\begin{array}{c|c||}
\hline
g& {\rm E}_{(7,3)}\\\hline
 1.3 & 12.0182 \\
 1.4 & 12.3149 \\
 1.5 & 12.6083 \\
 1.6 & 12.8979 \\
 1.7 & 13.1833 \\
 1.8 & 13.4643 \\
 1.9 & 13.7407 \\
 2. & 14.0124 \\
 2.1 & 14.2796 \\
 2.2 & 14.5423 \\
 2.3 & 14.8006 \\
 2.4 & 15.0546 \\
 2.5 & 15.3046
  \\\hline
\end{array}
\begin{array}{c|c||}
\hline
g& {\rm E}_{(7,3)}\\\hline
 2.6 & 15.5507 \\
 2.7 & 15.7929 \\
 2.8 & 16.0315 \\
 2.9 & 16.2667 \\
 3. & 16.4984 \\
 3.1 & 16.727 \\
 3.2 & 16.9524 \\
 3.3 & 17.1748 \\
 3.4 & 17.3944 \\
 3.5 & 17.6112 \\
 3.6 & 17.8253 \\
 3.7 & 18.0369 \\
 3.8 & 18.2459
  \\\hline
\end{array}
\begin{array}{c|c||}
\hline
g& {\rm E}_{(7,3)}\\\hline
 3.9 & 18.4526 \\
 4. & 18.6569 \\
 4.1 & 18.859 \\
 4.2 & 19.059 \\
 4.3 & 19.2568 \\
 4.4 & 19.4525 \\
 4.5 & 19.6463 \\
 4.6 & 19.8382 \\
 4.7 & 20.0282 \\
 4.8 & 20.2164 \\
 4.9 & 20.4029 \\
 5. & 20.5876 \\
 5.1 & 20.7707
  \\\hline
\end{array}
\begin{array}{c|c||}
\hline
g& {\rm E}_{(7,3)}\\\hline
 5.2 & 20.9521 \\
 5.3 & 21.132 \\
 5.4 & 21.3103 \\
 5.5 & 21.4871 \\
 5.6 & 21.6625 \\
 5.7 & 21.8365 \\
 5.8 & 22.009 \\
 5.9 & 22.1802 \\
 6. & 22.3501 \\
 6.1 & 22.5187 \\
 6.2 & 22.686 \\
 6.3 & 22.852 \\
 6.4 & 23.0169
  \\\hline
\end{array}
\begin{array}{c|c|}
\hline
g& {\rm E}_{(7,3)}\\\hline
 6.5 & 23.1806 \\
 6.6 & 23.3431 \\
 6.7 & 23.5045 \\
 6.8 & 23.6647 \\
 6.9 & 23.8239 \\
 7. & 23.9821 \\
 7.1 & 24.1391 \\
 7.2 & 24.2952 \\
 7.3 & 24.4502 \\
 7.4 & 24.6043 \\
 7.5 & 24.7574 \\
 7.6 & 24.9095 \\
 7.7 & 25.0607
  \\\hline
\end{array}
\eea
}

\subsection*{$J=8,n=4$ operator}

{\smaller \bea\la{E84data}
\begin{array}{|c|c||}
\hline
g& {\rm E}_{(8,4)} \\\hline
0& 10 \\
0.1 & 10.0384 \\
 0.2 & 10.1495 \\
 0.3 & 10.3227 \\
 0.4 & 10.5452 \\
 0.5 & 10.8049 \\
 0.6 & 11.0922 \\
 0.7 & 11.3996 \\
 0.8 & 11.7217 \\
 0.9 & 12.0542 \\
 1. & 12.3942 \\
 1.1 & 12.739 \\
 1.2 & 13.0867  \\\hline
\end{array}
\begin{array}{c|c||}
\hline
g& {\rm E}_{(8,4)} \\\hline
 1.3 & 13.4355 \\
 1.4 & 13.7839 \\
 1.5 & 14.1306 \\
 1.6 & 14.4743 \\
 1.7 & 14.814 \\
 1.8 & 15.1491 \\
 1.9 & 15.4789 \\
 2. & 15.8032 \\
 2.1 & 16.1218 \\
 2.2 & 16.4347 \\
 2.3 & 16.742 \\
 2.4 & 17.0438 \\
 2.5 & 17.3404  \\\hline
\end{array}
\begin{array}{c|c||}
\hline
g& {\rm E}_{(8,4)} \\\hline
 2.6 & 17.6319 \\
 2.7 & 17.9186 \\
 2.8 & 18.2007 \\
 2.9 & 18.4784 \\
 3. & 18.7518 \\
 3.1 & 19.0213 \\
 3.2 & 19.2868 \\
 3.3 & 19.5486 \\
 3.4 & 19.8068 \\
 3.5 & 20.0617 \\
 3.6 & 20.3132 \\
 3.7 & 20.5616 \\
 3.8 & 20.8069  \\\hline
\end{array}
\begin{array}{c|c||}
\hline
g& {\rm E}_{(8,4)} \\\hline
 3.9 & 21.0493 \\
 4. & 21.2889 \\
 4.1 & 21.5257 \\
 4.2 & 21.7598 \\
 4.3 & 21.9915 \\
 4.4 & 22.2206 \\
 4.5 & 22.4474 \\
 4.6 & 22.6718 \\
 4.7 & 22.894 \\
 4.8 & 23.114 \\
 4.9 & 23.3318 \\
 5. & 23.5476 \\
 5.1 & 23.7615  \\\hline
\end{array}
\begin{array}{c|c||}
\hline
g& {\rm E}_{(8,4)} \\\hline
 5.2 & 23.9733 \\
 5.3 & 24.1833 \\
 5.4 & 24.3914 \\
 5.5 & 24.5978 \\
 5.6 & 24.8024 \\
 5.7 & 25.0053 \\
 5.8 & 25.2065 \\
 5.9 & 25.4061 \\
 6. & 25.6041 \\
 6.1 & 25.8006 \\
 6.2 & 25.9955 \\
 6.3 & 26.189 \\
 6.4 & 26.3811 \\\hline
\end{array}
\begin{array}{c|c||}
\hline
g& {\rm E}_{(8,4)} \\\hline
 6.5 & 26.5717 \\
 6.6 & 26.761 \\
 6.7 & 26.9489 \\
 6.8 & 27.1355 \\
 6.9 & 27.3208 \\
 7. & 27.5049 \\
 7.1 & 27.6877 \\
 7.2 & 27.8693 \\
 7.3 & 28.0497 \\
 7.4 & 28.2289 \\
 7.5 & 28.407 \\
 7.6 & 28.584 \\
 7.7 & 28.7598  \\\hline
\end{array}
\eea
}



\begin{thebibliography}{20}


\bibitem{M}
  J.~M.~Maldacena,
  ``The large N limit of superconformal field theories and supergravity,''
  Adv.\ Theor.\ Math.\ Phys.\  {\bf 2} (1998) 231
  [Int.\ J.\ Theor.\ Phys.\  {\bf 38} (1999) 1113]
  [arXiv:hep-th/9711200].

 \bibitem{Zamolodchikov90}
  A.~B.~Zamolodchikov,
  ``Thermodynamic Bethe Ansatz in Relativistic Models. Scaling Three State Potts and Lee--Yang Models,''
 {\slshape   Nucl.\ Phys.\  B }{\bf 342} (1990) 695.

\bibitem{DT96}
  P.~Dorey and R.~Tateo,
  ``Excited states by analytic continuation of TBA equations,''
 {\slshape   Nucl.\ Phys.\  B }{\bf 482} (1996) 639
  [arXiv:hep-th/9607167].

\bibitem{BLZe}
  V.~V.~Bazhanov, S.~L.~Lukyanov and A.~B.~Zamolodchikov,
  ``Quantum field theories in finite volume: Excited state energies,''
  Nucl.\ Phys.\  B {\bf 489} (1997) 487, hep-th/9607099.

\bibitem{AF07}
  G.~Arutyunov and S.~Frolov,
  ``On String S-matrix, Bound States and TBA,''
  JHEP {\bf 0712} (2007) 024, hep-th/0710.1568.

 \bibitem{BS}
  N.~Beisert and M.~Staudacher,
  ``Long-range $PSU(2,2|4)$ Bethe ansaetze for gauge theory and strings,''
 Nucl.\ Phys.\  B {\bf 727} (2005) 1
  [arXiv:hep-th/0504190].

  N.~Beisert, B.~Eden and M.~Staudacher,
  ``Transcendentality and crossing,''
  J.\ Stat.\  Mech.\  {\bf 0701} (2007) P021
  [arXiv:hep-th/0610251].



\bibitem{AF09a}
  G.~Arutyunov and S.~Frolov,
  ``String hypothesis for the $\AdS$ mirror,''
  JHEP {\bf 0903} (2009) 152
  [arXiv:0901.1417 [hep-th]].


 \bibitem{AF09b}
  G.~Arutyunov and S.~Frolov,
  ``Thermodynamic Bethe Ansatz for the $\AdS$ Mirror Model,''
  JHEP {\bf 0905} (2009) 068
  [arXiv:0903.0141]. $\bullet$  ``Simplified TBA equations of the $\AdS$ mirror model,''
  JHEP {\bf 0911} (2009) 019
  [arXiv:0907.2647].

\bibitem{BFT}
  D.~Bombardelli, D.~Fioravanti and R.~Tateo,
  ``Thermodynamic Bethe Ansatz for planar AdS/CFT: a proposal,''
  J.\ Phys.\ A  {\bf 42} (2009) 375401
  [arXiv:0902.3930].

\bibitem{GKKV09}
  N.~Gromov, V.~Kazakov, A.~Kozak and P.~Vieira,
  ``Exact Spectrum of Anomalous Dimensions of Planar N = 4 Supersymmetric
  Yang-Mills Theory: TBA and excited states,''
  Lett.\ Math.\ Phys.\  {\bf 91} (2010) 265
  [arXiv:0902.4458 [hep-th]].
  
\bibitem{AFS09}
  G.~Arutyunov, S.~Frolov and R.~Suzuki,
  ``Exploring the mirror TBA,''
  JHEP {\bf 1005} (2010) 031
  [arXiv:0911.2224 [hep-th]].

\bibitem{BH10b}
  J.~Balog, A.~Hegedus,
  ``The Bajnok-Janik formula and wrapping corrections,''
  JHEP {\bf 1009}, 107 (2010).
  [arXiv:1003.4303 [hep-th]].
  
\bibitem{Sfondrini:2011rr}
  A.~Sfondrini, S.~J.~van Tongeren,
  ``Lifting asymptotic degeneracies with the Mirror TBA,''
  JHEP {\bf 1109 } (2011)  050.
  [arXiv:1106.3909 [hep-th]].


\bibitem{AFT11}
  G.~Arutyunov, S.~Frolov and S.~J.~van Tongeren,
  ``Bound States in the Mirror TBA,''
  arXiv:1111.0564 [hep-th].

\bibitem{KP}
A. Kl\"umper, P. A. Pearce,  ``Conformal weights of RSOS lattice
models and their fusion hierarchy,''  Physica A {\bf 183} (1992) 304-350. $\bullet$
 ``New results for exactly
solvable critical RSOS models and vertex models in two dimensions,'' Physica A {\bf 194} (1993) 397-405.

\bibitem{Arutyunov:2011uz}
  G.~Arutyunov and S.~Frolov,
  ``Comments on the Mirror TBA,''
  JHEP {\bf 1105} (2011) 082
  [arXiv:1103.2708 [hep-th]].
 
 
\bibitem{Gromov09a}
 N.~Gromov,
  ``Y-system and Quasi-Classical Strings,''
  JHEP {\bf 1001} (2010) 112
  [arXiv:0910.3608 [hep-th]].
 
 
 
   \bibitem{AFS10}
   G.~Arutyunov, S.~Frolov and R.~Suzuki,
  ``Five-loop Konishi from the Mirror TBA,''
  JHEP {\bf 1004} (2010) 069
  [arXiv:1002.1711 [hep-th]].

\bibitem{BH10a}
 J.~Balog and A.~Hegedus,
  ``5-loop Konishi from linearized TBA and the XXX magnet,''
  JHEP {\bf 1006} (2010) 080
  [arXiv:1002.4142 [hep-th]].


 \bibitem{BJ08}
  Z.~Bajnok and R.~A.~Janik,
  ``Four-loop perturbative Konishi from strings and finite size effects for multiparticle states,''
 Nucl.\ Phys.\  B {\bf 807} (2009) 625
  [arXiv:0807.0399].

  \bibitem{BJ09}
   Z.~Bajnok, A.~Hegedus, R.~A.~Janik and T.~Lukowski,
  ``Five loop Konishi from AdS/CFT,''
  Nucl.\ Phys.\  B {\bf 827} (2010) 426
  [arXiv:0906.4062 [hep-th]].


\bibitem{LRV09}
 T.~Lukowski, A.~Rej and V.~N.~Velizhanin,
  ``Five-Loop Anomalous Dimension of Twist-Two Operators,''
  Nucl.\ Phys.\  B {\bf 831} (2010) 105
  [arXiv:0912.1624 [hep-th]].

\bibitem{Janik:2010kd}
  R.~A.~Janik,
  ``Review of AdS/CFT Integrability, Chapter III.5: Luscher corrections,''
  arXiv:1012.3994 [hep-th].


\bibitem{Sieg}
  F.~Fiamberti, A.~Santambrogio, C.~Sieg and D.~Zanon,
  ``Wrapping at four loops in N=4 SYM,''
  Phys.\ Lett.\  B {\bf 666} (2008) 100
  [arXiv:0712.3522 [hep-th]].

\bibitem{Velizhanin:2008jd}
  V.~N.~Velizhanin,
  ``The four-loop anomalous dimension of the Konishi operator in N=4 supersymmetric Yang-Mills theory,''
  JETP Lett.\  {\bf 89 } (2009)  6-9.
  [arXiv:0808.3832 [hep-th]].


  

\bibitem{GKV09b}
 N.~Gromov, V.~Kazakov and P.~Vieira,
  ``Exact Spectrum of Planar ${\cal N}=4$ Supersymmetric Yang-Mills Theory:
  Konishi Dimension at Any Coupling,''
  Phys.\ Rev.\ Lett.\  {\bf 104} (2010) 211601
  [arXiv:0906.4240 [hep-th]].

\bibitem{F10}
  S.~Frolov,
  ``Konishi operator at intermediate coupling,''
  J.\ Phys.\ A  {\bf 44} (2011) 065401
  [arXiv:1006.5032 [hep-th]].

\bibitem{Gromov:2011de}
  N.~Gromov, D.~Serban, I.~Shenderovich, D.~Volin,
  ``Quantum folded string and integrability: From finite size effects to Konishi dimension,''
  JHEP {\bf 1108 } (2011)  046.
  [arXiv:1102.1040 [hep-th]].

\bibitem{Roiban:2011fe}
  R.~Roiban, A.~A.~Tseytlin,
  ``Semiclassical string computation of strong-coupling corrections to dimensions of operators in Konishi multiplet,''
  Nucl.\ Phys.\  {\bf B848 } (2011)  251-267.
  [arXiv:1102.1209 [hep-th]].

\bibitem{Vallilo:2011fj}
  B.~C.~Vallilo, L.~Mazzucato,
  ``The Konishi multiplet at strong coupling,''
  [arXiv:1102.1219 [hep-th]].

\bibitem{Beccaria:2011uz}
  M.~Beccaria, G.~Macorini,
  ``Quantum folded string in $S^5$ and the Konishi multiplet at strong coupling,''
  [arXiv:1108.3480 [hep-th]].

\bibitem{AFS}
  G.~Arutyunov, S.~Frolov and M.~Staudacher,
  ``Bethe ansatz for quantum strings,''
  JHEP {\bf 0410} (2004) 016
  [arXiv:hep-th/0406256].

\bibitem{GKV02}
  S.~S.~Gubser, I.~R.~Klebanov and A.~M.~Polyakov,
  ``Gauge theory correlators from non-critical string theory,''
  Phys.\ Lett.\  B {\bf 428} (1998) 105
  [arXiv:hep-th/9802109].
  
\bibitem{AF05}
  G.~Arutyunov and S.~Frolov,
  ``Uniform light-cone gauge for strings in $\AdS$: Solving $\su(1|1)$
  sector,''
  JHEP {\bf 0601} (2006) 055
  [arXiv:hep-th/0510208].

\bibitem{Plefka}
  F.~Passerini, J.~Plefka, G.~W.~Semenoff and D.~Young,
  ``On the Spectrum of the $\AdS$ String at large lambda,''
  JHEP\ {\bf 1103} (2011) 046
  [arXiv:1012.4471 [hep-th]].

\bibitem{RT}
  R.~Roiban and A.~A.~Tseytlin,
  ``Quantum strings in $\AdS$: strong-coupling corrections to dimension of Konishi operator,''
  JHEP {\bf 0911} (2009) 013
  [arXiv:0906.4294 [hep-th]].

\bibitem{Basso}
  B.~Basso,
  ``An exact slope for AdS/CFT,''
  arXiv:1109.3154 [hep-th].

\bibitem{Gr11}
  N.~Gromov and S.~Valatka,
  ``Deeper Look into Short Strings,''
  arXiv:1109.6305 [hep-th].





\end{thebibliography}
\end{document}